\documentclass[11pt]{article}
\usepackage{fullpage}
\RequirePackage{amsthm,amsmath,amsfonts,amssymb,graphicx,subcaption,multirow,bm,placeins,xcolor}
\usepackage[hidelinks]{hyperref}
\usepackage[round]{natbib}
\usepackage{authblk}

\title{\vspace{-1cm}Hierarchical Dynamic Modeling \\ for Individualized Bayesian Forecasting}
\author[1]{Anna K. Yanchenko}
\author[1]{Di Daniel Deng}
\author[2]{Jinglan Li}
\author[2]{Andrew J. Cron}
\author[1]{Mike West}

\affil[1]{Department of Statistical Science,
Duke University, Durham NC 27708-0251. U.S.A.}

\affil[2]{$84.51^\circ$, 100 West 5th Street, Cincinnati, OH 45202.  U.S.A.}
\date{\today}
\begin{document}
\maketitle

\begin{abstract}
We present a case study and methodological developments in large-scale hierarchical dynamic modeling for personalized prediction in commerce. 
The context is supermarket  sales, where improved forecasting of customer/household-specific purchasing behavior informs decisions about personalized pricing and promotions on a continuing basis. This is a big data, big modeling and forecasting setting involving many thousands of customers and items on sale, requiring sequential analysis,  addressing information flows at multiple levels over time, and with heterogeneity of customer profiles and item categories.  Models developed are fully Bayesian, interpretable  and multi-scale, with hierarchical forms overlaid on the inherent structure of the retail setting.  Customer behavior is modeled at several levels of aggregation,  and information  flows from aggregate to individual levels.
Forecasting at an individual  household level    infers price sensitivity  to inform personalized pricing and promotion decisions. Methodological innovations include extensions of Bayesian dynamic mixture models, their integration into multi-scale systems, and forecast evaluation with context-specific  metrics.  The use of simultaneous predictors from multiple hierarchical levels  improves forecasts at the customer-item level of main interest. This is evidenced across many different households and items, indicating the utility of the modeling framework for this and other individualized  forecasting applications.

\end{abstract}

\section{Introduction\label{sec:intro}}

Practical forecasting to inform decision making  in large retail and allied commercial demand/sales settings presents enormous challenges to statistics. Problems involve big, complex and heterogeneous data,  sparsity of information, and inherently hierarchical structures in multiple dimensions (time, geography, customers/households, and items on sale). Our work  focuses on such challenges in a supermarket sales setting; a primary goal is to forecast customer behavior at the individual (customer/household) and item (specific product  on sale) level, in each store in  a large national supermarket system, and continually over time.   Forecasts  inform downstream decision making  about prices and promotions.  The fine-scale decisions lie at the level of individual items potentially purchased each week by individual customers/households, involving assessments of price sensitivity so as to best customize/individualize offers of item-specific discounts and other marketing decisions.   

Analysis involves large data sets on past purchasing behavior of many thousands of households on a weekly basis. The data typically includes item purchase information by household, with details of item prices and promotions, for the many items in the retail chain and all customers. 
Purchasing behavior is hugely heterogeneous; many households will not purchase a specific item for multiple weeks if at all, making the data  sparse and   very variable across items.   Households and items are   nested within intersecting hierarchies,  and understanding purchasing trends at each level is central.    Sharing information across households  is a key interest, but is very challenging in terms of both computational demands and data heterogeneity. At a strategic level,  the following {\em desiderata} arise in the applied context;  models for personalized forecasting should be 
\begin{itemize}
\itemsep0pt
\item able to incorporate predictor information such as price and promotions, 
\item adaptable to time-varying trends, regression effects and unforeseen temporal changes, 
\item interpretable and open to intervention by users and downstream decision makers, 
\item fully probabilistic to properly characterize forecast uncertainties and allow formal model and forecast assessment under multiple metrics,  
\item adapted to hierarchical settings, and 
\item  amenable to automated, computationally efficient sequential learning and forecasting. 
\end{itemize}
Statistical challenges raised by these needs are shared by increasingly fine-scale prediction problems in various fields due to the  growing availability of individual level data.


\subsection*{Related and Relevant Work}
Our developments link to interests  in  individualized forecasting in a variety of retail settings~\citep[e.g.][]{8637442}, while the broader statistical challenges relate to personalization as a main goal of  recommendation systems in a variety of application areas.  Applications range from image recommendation systems~\citep{Niu:2018} to music~\citep{Wang:2013}. 
Personalized recommendation systems typically rely on methods such as collaborative filtering and matrix factorization~\citep{10.1155/2009/421425, Du:2018} or Bayesian personalized ranking~\citep{BPR}.  Recent approaches have leveraged deep learning to scale-up to larger and more complex data~\citep{He:2018, Niu:2018, DBLP:journals/corr/abs-1906-00091}.  The majority of such approaches are inherently non-dynamic, though matrix factorization has been extended to a dynamic setting~\citep{pmlr-v54-jerfel17a} and temporal features can be leveraged~\citep{10.1145/1526709.1526802, 7195574}.  Recommendation system approaches typically aim to match users to items, rather than aiming to forecast which items an individual will purchase.
In medical applications,  a core focus on personalized prediction arose with the advent of  genomics~\citep[e.g.][]{Nevins2003,Pittman2004,West2006}; the field has grown and  seen advances based on statistics and machine learning in areas such as 
Alzheimer's~\citep{DBLP:journals/corr/abs-1712-00181, Fisher:2019} and glaucoma progression prediction~\citep{Kazemian:2018}. 
Such approaches have exploited various techniques, including traditional Kalman filtering~\citep{Kazemian:2018}, Gaussian processes~\citep{DBLP:journals/corr/abs-1712-00181} and variants of Boltzmann machines~\citep{Fisher:2019}.  The goal is usually forecasting a single quantity of interest (i.e. glaucoma progression) rather than forecasting across related data ``dimensions'' such  as households and items.  
 
In the retail domain, individualized modeling  has generally involved either 
 (i) traditional random effects models, which are often not explicitly dynamic and can be challenging to fit to many time series, or (ii) black-box machine learning approaches that generally lack  interpretability and  probabilistic structure.  
Random effects models~\citep{Lichman:2018, Kazemian:2018, Lichman:2018, 5992380}  are inherently hierarchical, but standard formulations do not meet the {\em desiderata} for dynamics and computational scalability.  Machine learning approaches such as~\citet{sen2019think} have utilized global matrix factorization for information sharing across time series 
with temporal convolution networks. \citet{NIPS2019_8907} used Gaussian copula processes with a focus on retail forecasting,~\citet{DeepAR:2017} and~\citet{pmlr-v97-wang19k} proposed deep models with probabilistic forecasts, while~\citet{8637442}  used recurrent neural networks for predicting next-arrival times of customers.  These deep learning approaches are sequential and can be probabilistic~\citep{NIPS2019_8907, 8637442} as well as computationally efficient, but lack  interpretability and openness to communication and intervention by downstream decision makers. For example, it is unclear how to interpret the impact of price discounts on purchasing behavior, making it very challenging to determine which household to send discount coupons to for which items-- a primary driver of the interest in personalized models.    

\subsection*{Modeling Perspective and Framework}

We address the {\em desiderata} laid out above  in an holistic Bayesian forecasting framework  using tried-and-tested dynamic modeling components and concepts.  The hierarchical structure of the problem naturally invites extensions of multi-scale modeling ideas 
for efficient and effective sharing of information.  We build on extensions of dynamic generalized linear models~(DGLMs:~\citealp[][chap. 14]{West-Harrison}) to forecast individual household purchasing behavior.  DGLMs and related dynamic Bayesian time series models have been widely applied to much success, and recent extensions have focused on tailoring these approaches to count-valued time series~\citep{BerryWest2018DCMM,BerryWest2018DBCM} and on increasing computational   efficiency in hierarchical multivariate settings~\citep{LavineCronWest2020factorDGLMs}. 
Such developments have been partly motivated by retail forecasting applications at a more aggregate level.  
\citet{BerryWest2018DCMM} introduced the dynamic count mixture model (DCMM), a mixture of Bernoulli and Poisson DGLMs to model count time series with a high proportion of zeros for retail demand forecasting.   These models can also account for over-dispersion in the Poisson DGLM and, importantly, leverage aggregate information via a multi-scale modeling approach. \citet{BerryWest2018DCMM}, and the extensions to joint transaction and sales forecasting in~\citet{BerryWest2018DBCM},  have demonstrated the value of multi-scale modeling in a  variety of settings as well as the competitive forecasting performance of  linked systems of Bayesian DGLMs.    

We extend the multi-scale approach to hierarchically compose sets of models on different aspects of household consumer behavior, while maintaining computational efficiency.  
This involves a  broad perspective on hierarchical, multi-scale modeling to share information across both items and households at the individual level; fusing tiered sets of conditional models at different levels of the hierarchy, this can yield improved forecasting for individual households.  Methodological innovation also includes a new class of dynamic linear mixture models (DLMMs) for mixed binary/continuous time series, relevant for components of the multi-scale system.  Importantly, computational tractability and efficiency are at the forefront in methodological developments.  
All model components are interpretable, fully Bayesian/probabilistic, and defined for sequential learning and forecasting, making the approach ideally suited to general personalized forecasting applications.

Section~\ref{sec:data} discusses data and application goals, followed by  models and computational details in Section~\ref{sec:models}.  Section~\ref{sec:metrics} discusses  decision analytic perspectives on metrics and evaluation methods, and Section~\ref{sec:results} presents some  empirical forecast accuracy results at both aggregate and individual levels.  Concluding comments appear in Section~\ref{sec:conc}.

\section{Data Structure and Application Goals\label{sec:data}}
 
\subsection{Data Description}
Data informs on individual household purchasing histories in a large supermarket system.  Data records selected for this case study provide information on the weekly purchasing behavior of over 500{,}000 households across 112 weeks for over 200 unique items on sale.  A ``household'' is considered an ``individual'', and primary interest lies in forecasting weekly household purchasing trends.  Available covariates   include the total \$US spend for each household each week, the number and types of items purchased and the discount percentage (discount amount divided by the regular price), if applicable.  Different households can be offered different discounts for the same item in the same week, due to coupons and e-marketing, for example.  

\subsubsection{Selection of Households}

\begin{figure}[h!]
\centering
\begin{subfigure}[b]{0.9\textwidth}
   \includegraphics[width=1\linewidth]{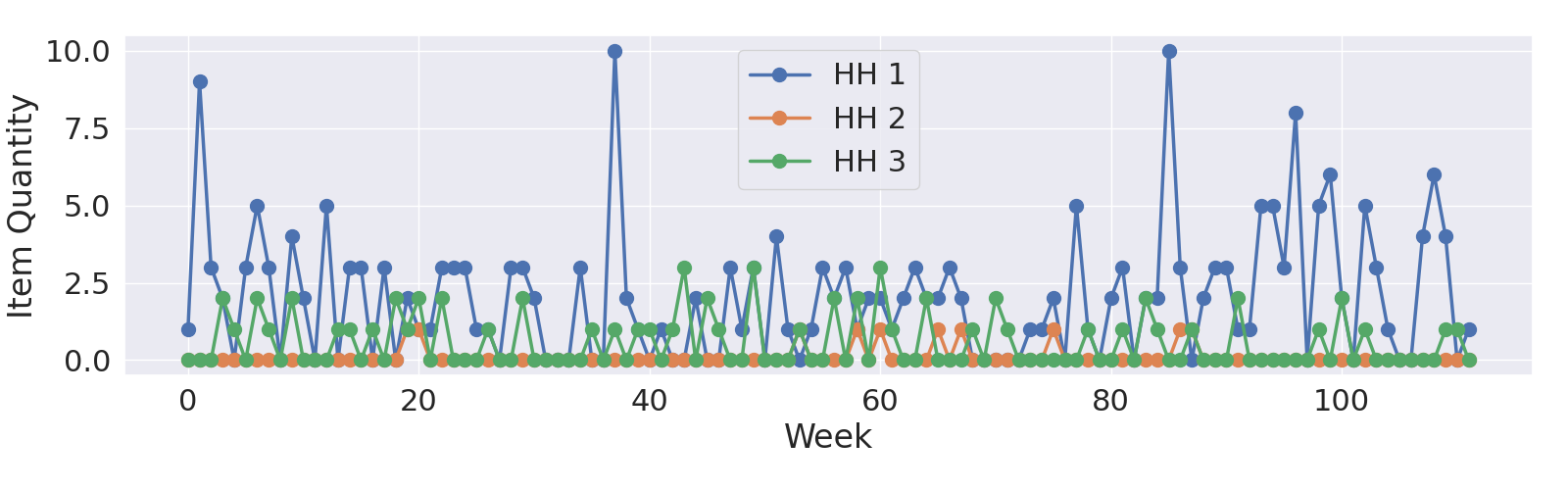}
   \caption{Household Group 1: High spending households.}
  \label{fig:Ng1a} 
\end{subfigure}
\begin{subfigure}[b]{0.9\textwidth}
   \includegraphics[width=1\linewidth]{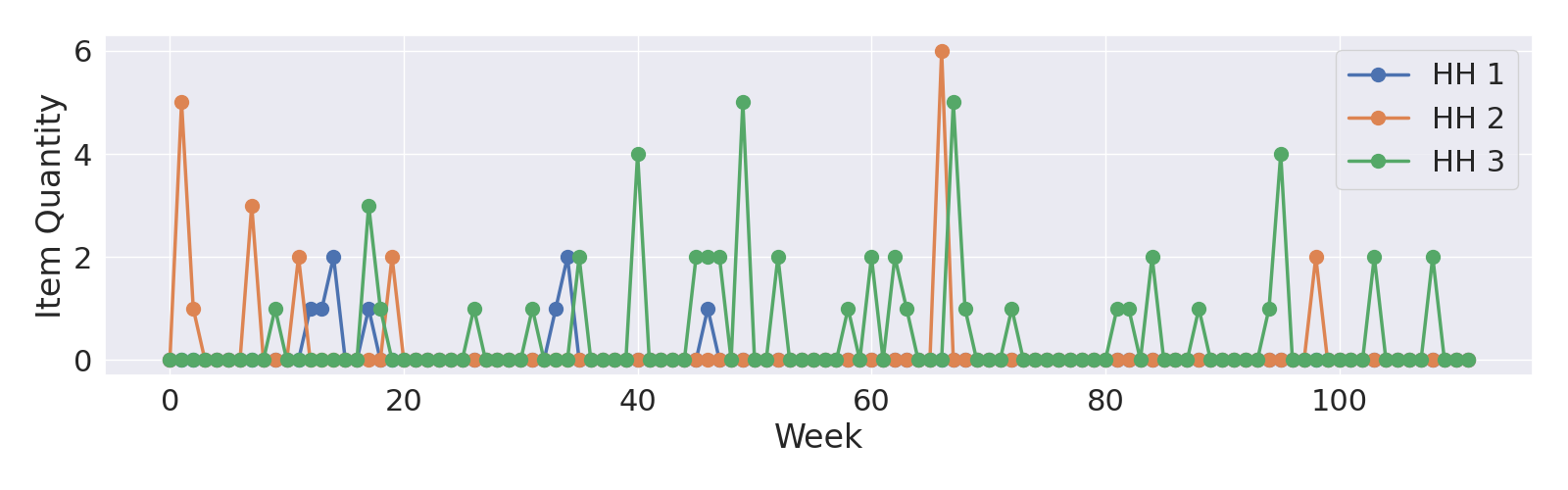}
   \caption{Household Group 3: Low spending households.}
  \label{fig:Ng2a}
\end{subfigure}
\caption{Purchasing trends for three different households in the (a) high-spending household group (Group 1) and (b) the low spending household group (Group 3) for one specific item.}\label{fig:HHs-1item}
\end{figure}

General buying trends differ considerably by household (\autoref{fig:HHs-1item},~\autoref{tab:EDA}) and there is additional industry interest in categorizing households by various demographic information and purchasing trends. We define three household groups based on total items  purchased over the course of the 112 weeks:
\begin{itemize}
\itemsep0pt
\item Household Group 1: high spending and purchasing households 
\item Household Group 2: moderate spending and purchasing households 
\item Household Group 3: lower spending and purchasing households 
\end{itemize}
Each household group  consists of 2{,}000 households; the methodology is scalable to many more households.  Additional demographic information can be incorporated into models or  groupings for evaluation, but is not available here.  The household groupings are used to explore forecast accuracy across a variety of households and to demonstrate that purchasing trends do vary significantly by household.  
\begin{table}[htp!]
\caption{Household group summaries.  Proportion of weeks that each household returns  with summary spend information.   There is variation in patterns of return and  total spend habits of households across groups.}
\vspace{-0.5cm}
\begin{center}
\scalebox{0.9}{\begin{tabular}{c|c|c|c|c}
Household Group	& Proportion Return  & 	Mean Spend &	Median Spend	& SD Spend \\\hline
1 &	0.96 &	\$ 29.48 & \$	26.12	& \$ 20.00 \\
2 &	0.94 &	\$ 19.23 &	 \$ 17.35	& \$ 12.05 \\
3 &	0.84 &	\$ 12.16 &	\$ 10.46	& \$ 8.38 \\
\end{tabular}}
\end{center}
\label{tab:EDA}
\end{table}%
\vspace{-0.75cm}

\subsubsection{Selection of Items}

\begin{figure}[b!]
\centering
\begin{subfigure}[b]{0.9\textwidth}
   \includegraphics[width=1\linewidth]{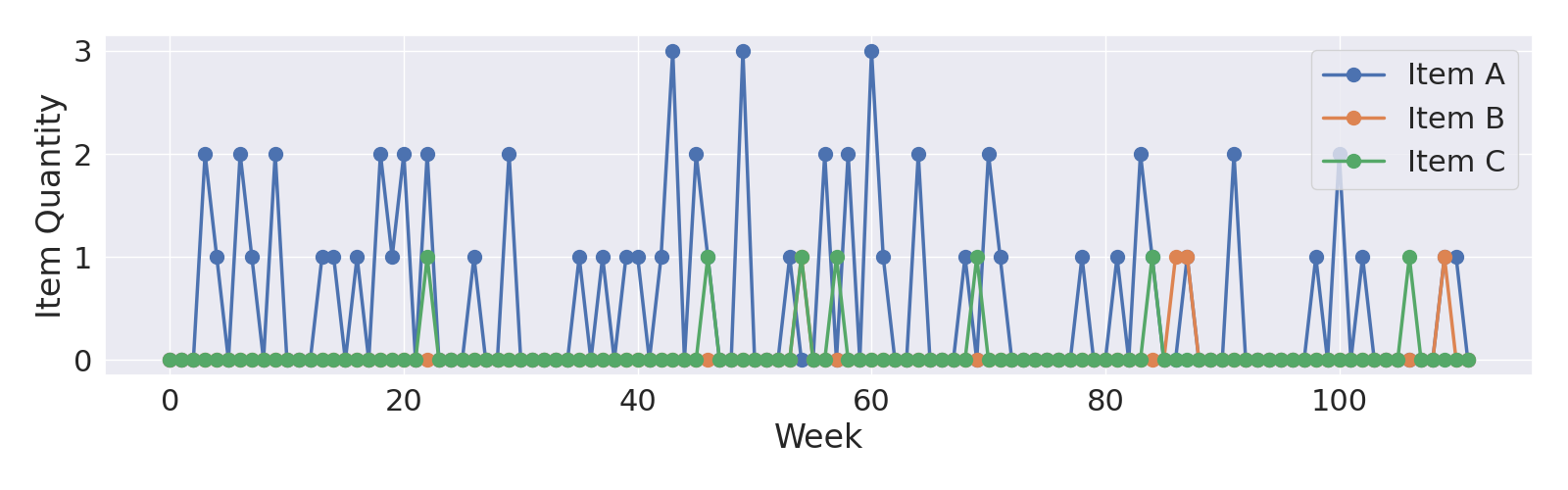}
   \caption{Household Group 1: High spending household.}
  \label{fig:Ng1b} 
\end{subfigure}
\begin{subfigure}[b]{0.9\textwidth}
   \includegraphics[width=1\linewidth]{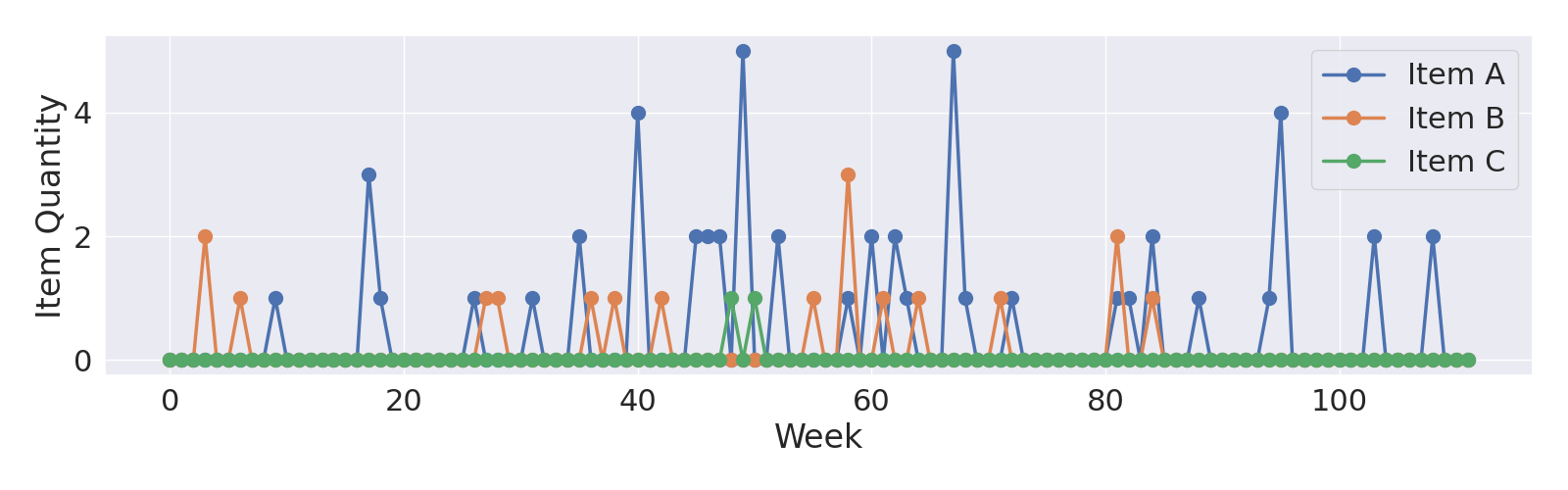}
   \caption{Household Group 3: Low spending household.}
  \label{fig:Ng2b}
\end{subfigure}
\caption{Purchasing trends for three different items for one individual household in the (a) high-spending household group (Group 1) and (b) the low spending household group (Group 3).  Purchasing trends vary greatly by item for a specific household, and again, many items are not purchased for many weeks.}\label{fig:items-1HH}
\end{figure}

Purchasing trends also differ significantly by item for individual households (\autoref{fig:items-1HH}). In addition to forecasting at the household level, a main goal is to understand price sensitivity of households on each item.  We select six specific items to exemplify this--  items that have dynamic discounts by household over different weeks, and  thus have ``potential'' for price sensitivity.  That is, aggregated over all 112 weeks, if an increase in discount amount leads to an increase in amount purchased for a majority of households, then many households tend to be price-sensitive for that item (at least in aggregate).  We categorize households based on their promotion scenarios and purchasing behaviors, with particular interest in households that are sensitive to promotions.  We then select items that have a high number of households that are price-sensitive in aggregate, choosing the six items with the largest proportion of  price-sensitive households.  These six items  are purchased by a large number of households in each household group, making the items very relevant exemplars.  Item A, shown in~\autoref{fig:HHs-1item}, is the main item of focus for subsequent  results; it is the highest selling item of the six considered, while the additional items, B--F, are discussed in more detail in the Appendix.   

For many weeks, the  households  in~\autoref{fig:items-1HH} do not purchase any of the items considered. It is possible that these households did not return to the store, or  that they did not specifically purchase any of the three items considered during a visit.  We thus need to separate \lq\lq no return to store'' from \lq\lq non-purchase'' as a model component.  

\begin{figure}[t!]
\centering
\begin{subfigure}{.45\textwidth}
  \centering
  \includegraphics[width=\linewidth]{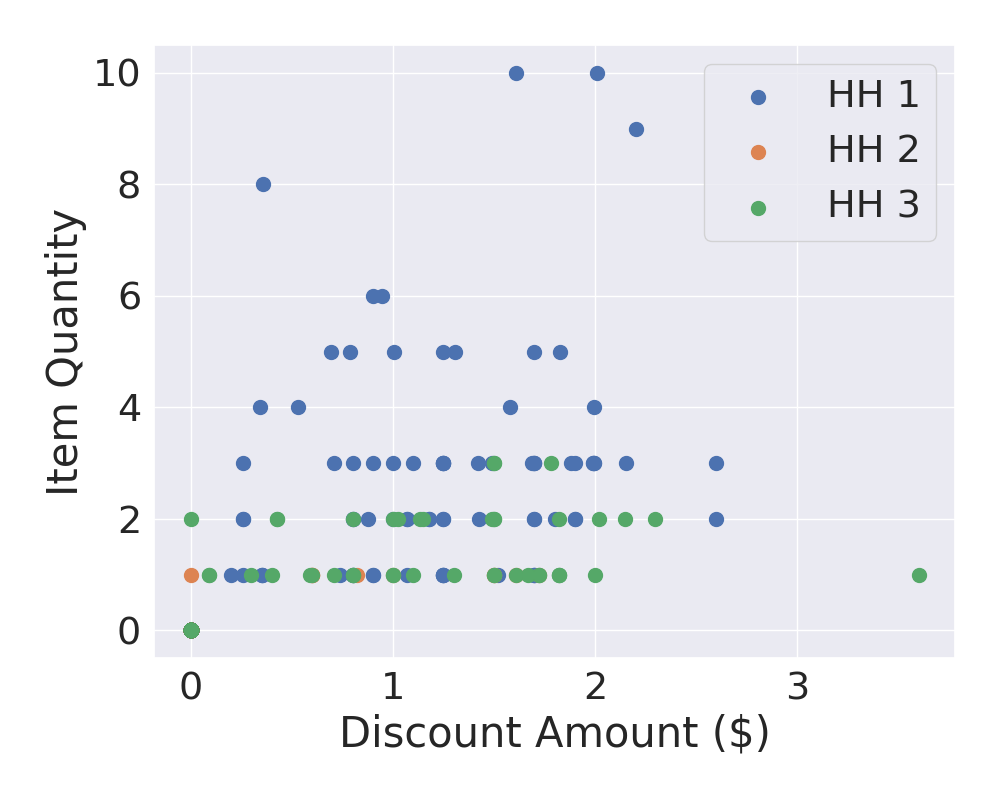}
  \caption{Household Group 1.}
 \label{fig:1a}
\end{subfigure}%
\begin{subfigure}{.45\textwidth}
  \centering
  \includegraphics[width=\linewidth]{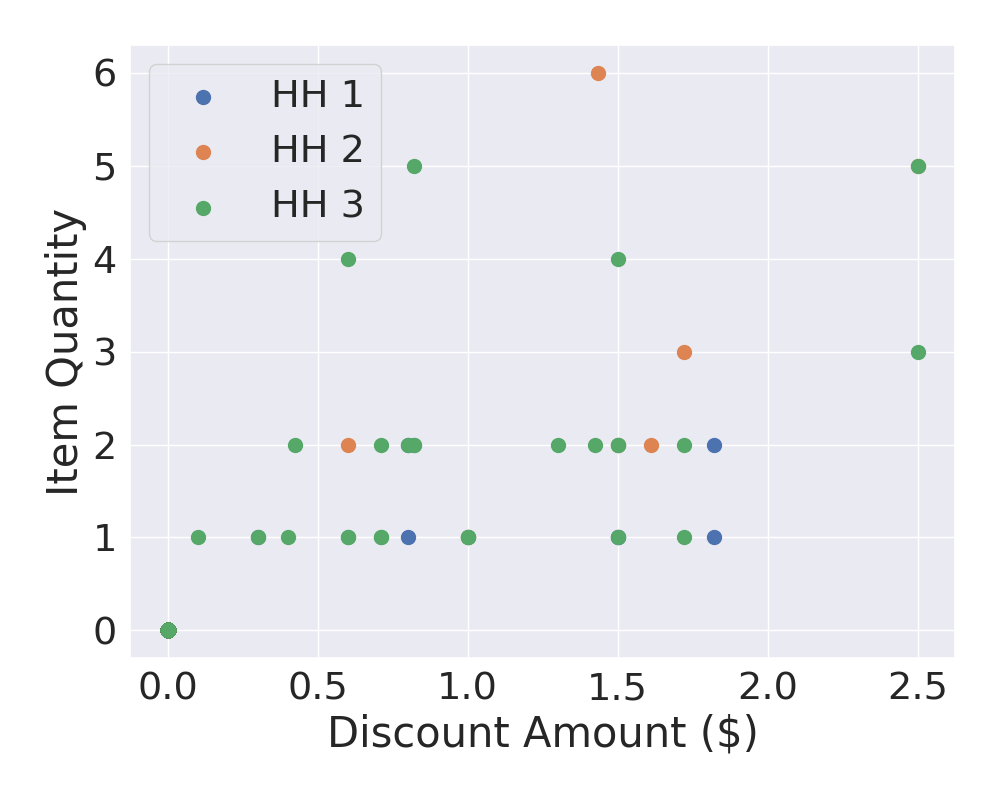}
  \caption{Household Group 3.}
 \label{fig:3a}
\end{subfigure}
\caption{Item quantity purchased by discount amount for Item A for three different households in household (a) Group 1 and (b) Group 3.  Each point represents an individual week.  Price sensitivity is heterogeneous by household, for a specific item.}
\label{fig:discount_vs_qty}
\end{figure}

Aggregate price sensitivity for selected items varies by household.  \autoref{fig:discount_vs_qty} displays  item quantity purchased vs.  discount amount offered for Item A across all weeks for three different households.  Some households (both higher spending and lower spending) purchase more of Item A when the discount amount is increased, while other households purchase the same item quantity, regardless of the discount amount offered.  Many households do not purchase the item, whatever the discount. This exemplifies  heterogeneity  by household in terms of price sensitivity, and analyzing this in a dynamic fashion is a main point of modeling focus.

\subsubsection{Nested Item Hierarchy}

\begin{figure}[hbt!]
\centering
\begin{subfigure}[b]{0.9\textwidth}
   \includegraphics[width=1\linewidth]{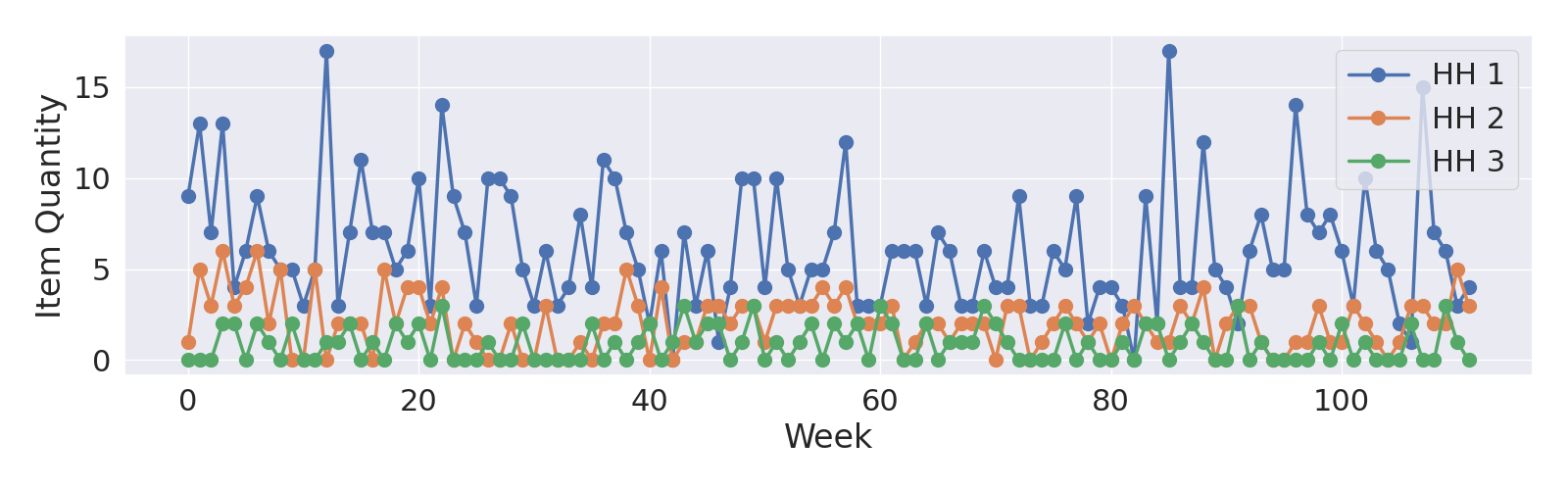}
   \caption{Household Group 1: Multiple Households.}
  \label{fig:Ng1c} 
\end{subfigure}
\begin{subfigure}[b]{0.9\textwidth}
   \includegraphics[width=1\linewidth]{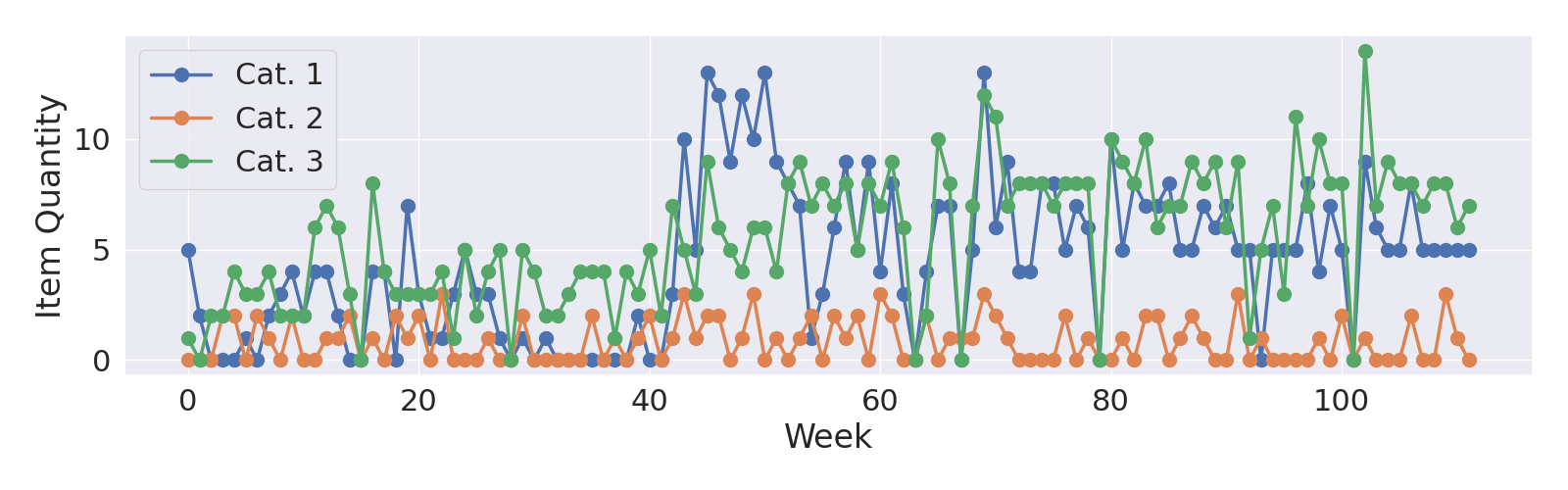}
   \caption{Household Group 1: Multiple Categories.}
  \label{fig:Ng2c}
\end{subfigure}
\caption{Purchasing trends for (a) three different households in the high-spending household group (Group 1) and (b) for one household in household Group 1 across three different categories. Purchasing trends are more stable and consistent at the category level.}
\label{fig:EDA-GRP2}
\end{figure}

Trends at the individual level tend to be very heterogeneous by household and item, and even week-to-week for a specific household-item pair.  However, there is a nested hierarchy of item information that provides opportunity to identify more stable trends at more aggregate levels.  The item hierarchy consists of category, sub-category and finally item information.  The category is the highest and broadest level of aggregation, for example a specific category might include all soda.  The sub-category level is more specific, and may include  diet sodas, for example.  Finally, the item level is the most specific, and would represent Diet Coke, for example. There are multiple items in each sub-category and multiple sub-categories in each  category.  
Purchasing trends tend to be much more stable and persistent when aggregated at the category level, especially for high spending households in household Group 1 (\autoref{fig:EDA-GRP2}).  Exploiting this aggregate information in modeling at the individual level will be important.

\subsection{Application Goals}

The overarching goal is to collectively forecast purchasing trends at the household-item level.  Several related sub-goals  inform our modeling perspective.   Understanding price sensitivity at an individual household level is especially important, as we aim to identify households that are likely to be sensitive to targeted discounts for specific items.  Related to this is the importance of probabilistic forecasting at all levels of the  hierarchy.   Individual household forecasts are intended for use within a larger forecasting and promotion  system,  requiring  probabilistic forecasts  for downstream decision making.  Then, due to the volume of data and the number of  households, computational efficiency is essential.   Methodology must scale to many hundreds of thousands of households and  items.


\section{Dynamic Modeling for Individualized Forecasting\label{sec:models}}

Our hierarchical modeling decomposition is multi-scale, allowing for the sharing of information without a need for complex dependency structures; forecast information from higher levels in the model informs lower-level forecasts. At each level, we use dynamic generalized linear models (DGLMs)~\citep{West-Harrison} and variants~\citep{BerryWest2018DCMM} to ensure that the overall modeling approach is sequential, interpretable, probabilistic and computationally efficient.  We use a new DGLM variant, the dynamic linear mixture model, to address key aspects of sparsity in  household data.

\subsection{Hierarchical Modeling Decomposition}

\begin{figure}[hbtp!]
\begin{center}
\includegraphics[width = 0.9\textwidth]{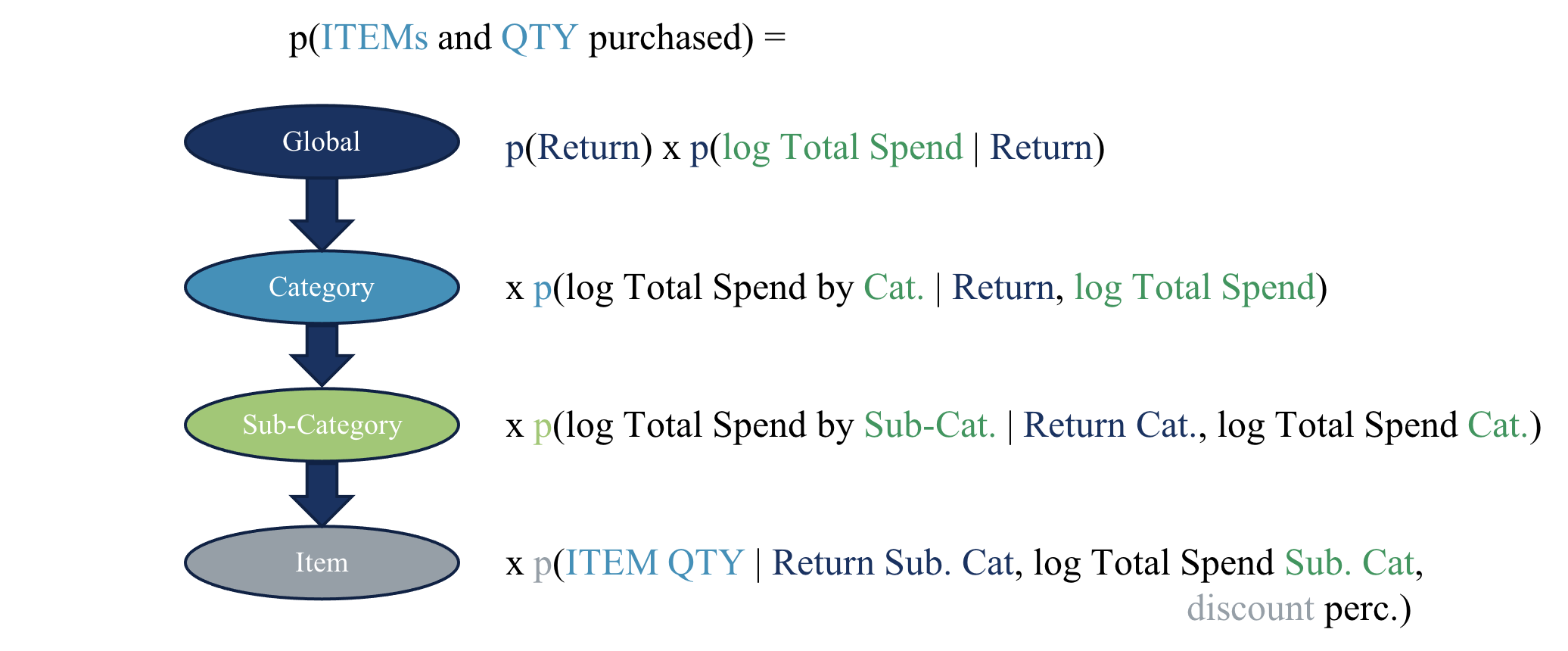}
\caption{Multi-scale model decomposition for one household-item pair.}
\label{fig:decomp}
\end{center}
\end{figure}
The approach to forecasting specific items that an individual household will purchase in a given week is motivated by concepts of rare-event modeling using nested sets of conditional probability models.  Trends at the individual household-item level tend to include many weeks of no purchase for the majority of items (\autoref{fig:HHs-1item}), making this a rare-event forecasting problem.  However, trends at the category-level, for example, in the nested item hierarchy tend to be more stable and persistent (\autoref{fig:EDA-GRP2}).  This concept applies also to sub-categories within categories, and then items within sub-categories. 
Hence the rare event of an individual household-item  purchase can be framed through conditional models of less rare events at the sub-category and then 
category levels. This multi-scale idea  conditions out ``no return to store'' in lower levels of the model and allows for the use of simultaneous information across levels. \autoref{fig:decomp} is a schematic of the main model structure for one household and one item. Here 
 \lq\lq Return'' is the indicator of whether the household shopped at all that week,  
(log) \lq\lq Total Spend'' represents the (log) total \$ amount spent by the household in that week (whether zero or positive) at global, category and sub-category levels, and \lq\lq discount perc.'' denotes the \% discount offered to the household on the nominal price of the item. 
At the target item level, we predict how many of that item the household purchases conditional on spending a non-zero \$ amount in the sub-category the item belongs to.  At this final stage, the outcome is a-- typically small-- non-negative integer. 

Each component of the model in~\autoref{fig:decomp} is dynamic across weeks and is fit to each household and item {\em conditionally} independently based on conditioning information.  Dependencies are captured through information sharing by  conditioning on aggregate information from higher levels in the hierarchy. 
This extends the multi-scale view that has been demonstrated to improve forecasting accuracy~\citep{BerryWest2018DCMM}.   The  decomposition also \lq\lq conditions out'' sparsity in the item level data, yielding improved forecast accuracy;  if we predict that a household does not return in a given week, we do not need to proceed further to the item level. 
Then-- importantly-- the decomposition allows for {\em simultaneous} predictors.  For example, after modeling the total spend in a given week, the forecast value of total spend can be used as a simultaneous predictor in the model for total spend at the category level for that week.  Such simultaneous predictors from  higher levels in the decomposition can greatly improve forecast accuracy in the lower-level models. Further, generating ensembles of higher-level quantities enables full probabilistic uncertainty propagation to finer levels.

\subsection{Dynamic Generalized Linear Models}

The building blocks at every level in~\autoref{fig:decomp} are based on DGLMs, so are 
fully probabilistic, naturally analyzed sequentially and with computational efficiency.  Coupled with the hierarchical decomposition into components at each level, this meets the {\em desiderata} of~Section~\ref{sec:intro}. 
 
In a general notation, let $y_t$ be a univariate time series observed at discrete times $t = 1, \ldots, T$.  Denote the available information at time $t$
by $\mathcal{D}_t = \{y_t, \mathcal{D}_{t-1}, \mathcal{I}_{t-1}\}$, where $\mathcal{I}_{t-1}$ represents any additional relevant information beyond the observed data.   Each $y_t$ follows a distribution in the exponential family with 
linear predictor $\lambda_t$ 
that evolves over time as 
\begin{equation}\label{eq:dglm}
\lambda_t = \bm F_t'\bm\theta_t \enspace\mbox{where}\enspace \bm\theta_t = \bm G_t\bm\theta_{t-1} + \bm\omega_t \enspace \mbox{and} \enspace \bm\omega_t\sim (\bm 0, \bm W_t),
\end{equation}
where
\begin{itemize}
\itemsep=0pt
 \item $\bm F_t$ is a matrix of known covariates at time $t$, 
 \item $\bm \theta_t$ is the state vector, which evolves via a first-order Markov process,
 \item $\bm G_t$ is a known state matrix,  
 \item $\bm \omega_t$ is the stochastic innovation vector, or evolution ``noise'',  with $\mathbb{E}(\bm \omega_t | \mathcal D_{t-1}, \mathcal I_{t-1}) = \bm 0$ and  $\mathbb{V}(\bm \omega_t | \mathcal D_{t-1}, \mathcal I_{t-1}) = \bm W_t,$  independently over time.
\end{itemize}
Model components in the hierarchy involve Bernoulli logistic DGLMs for p(Return), normal dynamic linear models (DLMs)-- with potentially time-varying conditional variances--  for p(log Total Spend $\vert$ Return), and Poisson loglinear DGLMs for item level counts. 

\subsection{Dynamic Count Mixture Models}

Many items will not be purchased in individual weeks even when a household spends on other items. Further, in some weeks a household will purchase more quantities than typical for a specific item.  Thus item quantity by household can exhibit both zero-inflation and over-dispersion.   Dynamic count mixture models (DCMMs) were introduced explicitly to address these non-Poisson (dynamic) features, and have very competitive forecasting accuracy compared to a variety of other models~\citep{BerryWest2018DCMM}. Hence, 
DCMMs are integrated into the hierarchical model structure at the final stage to predict the item quantity purchased by the specific household conditional on information from the earlier levels of the hierarchy together with item pricing and discount information. 
 
If $y_t$ is a non-negative count time series, let $z_t = \bm 1 (y_t > 0)$, where $\bm 1$ is the indicator function.  A DCMM has the form
\begin{equation}\label{eq:DCMM}
\begin{split}
z_t \sim\mbox{Bernoulli}\left(\pi_t\right)\; \mbox{and}\; (y_t |z_t) &= \begin{cases}
  0, & \text{if } z_t = 0, \\
  1 + s_t, \; s_t \sim\mbox{Poisson}\left(\mu_t\right), & \text{if } z_t = 1, 
\end{cases} \\
\textrm{with logit}(\pi_t) &= \bm F^{0'}_t\bm\xi_t \,\textrm{and}\, \log(\mu_t) = \bm F_t^{+'}\bm\theta_t,
\end{split}
\end{equation}
where $\bm\xi_t$ and $\bm\theta_t$ are state vectors following usual linear evolutions, and $\bm F_t^0$ and $\bm F_t^+$ are known regression vectors.  The DCMM models $\mu_t$ and $\pi_t$ independently, and $\bm F_t^0$ and $\bm F_t^+$ can be distinct, depending on the application.  Sequential learning and forecasting proceeds similarly as in the DGLM, with the Poisson model only being updated if $z_t = 1$; see the Appendix and full details in~\citet{BerryWest2018DCMM}.  

\subsection{Dynamic Linear Mixture Models}

At the category and sub-category levels there are also many zero values, indicating no purchase for that week.  This requires extensions of the traditional normal DLM for (log) \$ spend to account for weeks with zero spend.  To address this, we map the concept underlying the DCMM  to a new class of {\em dynamic linear mixture models (DLMMs)}.  This treats zero spend in a given week as a binary time series and, conditional on a non-zero spend, applies a DLM to log \$ amount spent.  The proportion of weeks a purchase is made by a household can vary greatly by category and sub-category, so the binary component here is key.  Some sub-categories, in particular, are purchased very infrequently and exhibit a high proportion of zeros, necessitating this new mixture modeling approach to capture conditionally (log) normally modeled outcomes combined with possibly many zeros.   DLMMs apply to the log total spend at both the category and sub-category levels, defining household-item specific forms for  p(Category log Total Spend $\vert$ Return, Global log Total Spend) and  p(Sub-Cat. log Total Spend $\vert$ Return Category, Category log Total Spend).  

Specifically, denote by $x_t$ a continuous valued outcome (log total spend at one level) and let $z_t = \bm 1\left(x_t \neq 0\right)$. A DLMM has the form  
\begin{equation}\label{eq:DLMM}
\begin{split}
z_t  \sim\mbox{Bernoulli}\left(\pi_t\right)\; \mbox{and}\; (x_t | z_t) &= \begin{cases}
  0, & \text{if } z_t = 0, \\
  x_t \sim\mathcal{N}\left(\bm F_t'\bm\theta_t, \; \bm V_t\right), & \text{if } z_t = 1,
\end{cases}
\end{split}
\end{equation}
where again $\pi_t$ is defined by a binary DGLM with its own state vector, and $\bm \theta_t$  follows a separate, independent DLM evolution and defines the model for non-zero \$ spend involving the known regression vectors $\bm F_t.$ 
As with the DCMM, these new DLMMs provide modeling flexibility in a fully Bayesian, computationally efficient extension of traditional DLMs; more details are noted in the Appendix.

\section{Forecast Evaluations and Metrics\label{sec:metrics}}

As highlighted in~\citet{BerryWest2018DCMM}, a forecast  is a full predictive distribution, as understanding uncertainty is critical for downstream decision making.  Any point forecasts used for model evaluation and comparison should be justified by an appropriate loss function in a Bayesian decision theoretic perspective.  Results can depend critically on loss functions selected so it is important to carefully consider which loss functions  are used to evaluate models.   

\subsection{Loss Functions and Point Forecasts}

Common choices of loss functions are squared error,  absolute deviation and  absolute percentage error, yielding mean or average values denoted by MSE, MAD and MAPE.  The latter is particularly common in commercial settings as it can putatively be compared across contexts, though is restricted to positive outcomes.  While forecast medians are optimal under MAD,  the optimal point forecast $f$ of $y\sim p(y)$ under MAPE is the $(-1)-$median, i.e.,  the median of $g(y)\propto p(y)/y$, always less than or equal to the median and sometimes much lower. 
 MAPE has been extended to the class of (zero-adjusted APE) ZAPE loss functions~\citep{Berry:2019,West2020decisionconstraints}. While there are several variants,  the form of ZAPE used here for a non-negative outcome $y$ and point forecast $f$ is
$\mathcal{L}_{ZAPE}(y, f) = \bm 1(y_t = 0) f/(1+f) + \bm 1(y_t > 0) | y-f|/y.  $
The ZAPE optimal forecast is always less than or equal to the $(-1)-$median; with forecast distributions heavily favoring zero or low values, it can often be zero. Numerical optimization for ZAPE forecasts is easy and detailed in the Appendix.

\subsection{Probabilistic Forecast Evaluations} 

We stress and routinely use calibration and coverage assessments for probabilistic forecasting, building on the increasing impact of full probabilistic assessment in commercial forecasting settings~\citep{BerryWest2018DCMM,BerryWest2018DBCM}.   For binary outcomes that are central in this application, frequency calibration is a core assessment concept.  To calculate realized calibration characteristics, we bin the probability scale and evaluate calibration within each probability bin.  In particular,  we consider calibration plots  related to  predicting when a household will return to shop, and how well we are modeling the sparsity in the data at several levels. 
Probability coverage comparing predicted versus realized  coverage over a  forecasting time period  uses predictive credible intervals (HPD, highest probability density), allowing assessments of under- or over-coverage across the probability scale via simple coverage plots.   This is used, for example, in assessment of components of the model forecasting household \$ spend at overall weekly basis and then within categories and sub-categories of items.

\section{Selected Summaries of Results\label{sec:results}}

Highlights summarized here focus on forecasting at the category, sub-category and item levels, with additional modeling results at all levels of the modeling decomposition given in the Appendix.

\subsection{Modeling Details}

We apply the hierarchical modeling decomposition of~\autoref{fig:decomp} to  households in each of the three groups (Group 1 = high spenders, Group 2 = moderate spenders, Group 3 = low spenders) to forecast item quantity purchased by household.  Each household is  modeled conditionally independently, but with information shared via conditioning variables-- across items and households at the item level-- as detailed in Section~\ref{subsec:multi}.  At the global level, p(Return) and p(log Total Spend $\vert$ Return) are modeled via a Bernoulli DGLM and a (log) normal DLM, respectively.  Here, we focus on results at (i) the category and sub-category levels using DLMMs, and (ii) the item level using DCMMs.  Unless otherwise noted, results are presented for Item A involving a high proportion of potentially price-sensitive households, and Item A's category and sub-category.  Additional items, sub-categories and categories are noted in the Appendix; results presented here generalize to a range of different items.   All models are implemented in the PyBats package~\citep{PyBats} with relevant extensions for the DLMM.   
%

In Sections~\ref{subsec:dlmm}--\ref{subsec:individ}, we treat simultaneous predictors from higher modeling levels as known.  This allows for an evaluation of specific aspects of each model at each modeling level.  However, the modeling decomposition is designed for fully simultaneous modeling, and in Section~\ref{subsec:simul} we use forecast values from higher levels in the modeling hierarchy as the simultaneous predictors, with only lagged predictors treated as known.  That is, the simultaneous predictors in the category, sub-category and item levels, as well as the prediction of return or not at all modeling levels, are the results of forecasts from the relevant higher modeling levels.  Additionally, while we focus on one-step ahead forecasts via simulation here, multi-step ahead forecasts also naturally proceed either via simulation or analytically~\citep{BerryWest2018DCMM,LavineCronWest2020factorDGLMs}.

Evaluations have included comparison with the DCMMs that have been found to be competitive in terms of forecasting accuracy when compared to a wide range of competing models~\citep{BerryWest2018DCMM}.  We find that our full modeling decomposition does improve forecasting results, essentially uniformly across examples studied, and regard this as substantial recommendation given the already proven utility of DCMMs.  There are, as far as we are aware, no other relevant approaches for comparison that satisfy all of the practically motivated desiderata:  interpretable models that are hierarchical and fully probabilistic,  analyzed and used sequentially, and parallelizable as well as inherently computationally scalable. 

\subsection{Simultaneous Predictors and DLMMs}\label{subsec:dlmm}

We have found that simultaneous predictors can contribute very substantially to forecast accuracy,  especially at lower levels in the modeling hierarchy. Predictors ``one level up'' in the modeling hierarchy are typically most useful and supercede information from higher levels. 
For example, predicted \$ spend at the sub-category level leads to improved forecasts at the item level relative to predictors based on spend at the category or global level.  This empirically reflects a natural conditional independence structure in the hierarchical decomposition. Specific examples  at the category, sub-category and item levels are now discussed.

\subsubsection{Category Level}

A DLMM defines  p(log Total Spend by Category $\vert$ Return, Global log Total Spend).  We compare models with 3 different predictors to evaluate the utility of the simultaneous predictors as compared to lagged predictors, with all covariates initially treated as known.  Denote the models by M1, M2, M3. All models have  a random-walk local level (trend term) and one model-specific dynamic regression term with predictor variable as follows: M1 has a lagged local predictor given as  the log total spend in the category at the last return;   M2 has a lagged global predictor given as the log total spend across all categories at the last return; M3 has a simultaneous global predictor given as the log total spend across all categories for the {\em current} week.  Aggregate results within each of the three household groups are given in~\autoref{tab:grp2-6}.  For each  metric, the point forecast is optimal under that loss function. MAPE is only evaluated at non-zero outcomes.  
 
\begin{figure}[h]
\centering
\scalebox{0.8}{
\begin{tabular}{cc}
 \includegraphics[width=.45\linewidth]{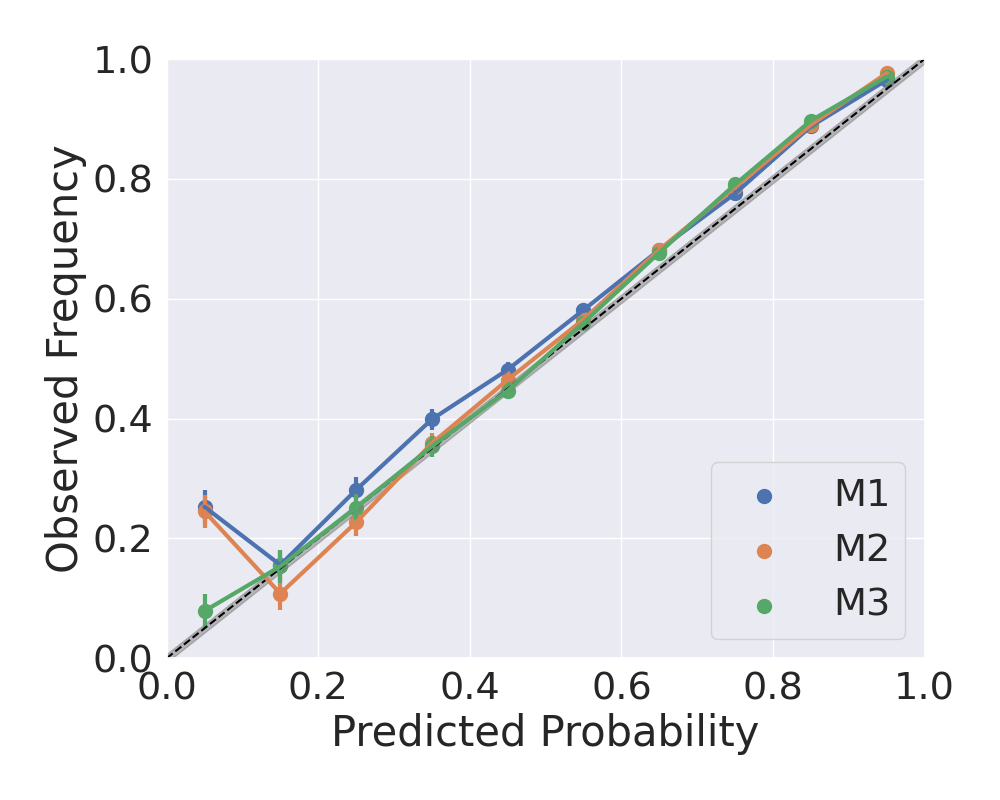} & 
 \includegraphics[width=.45\linewidth]{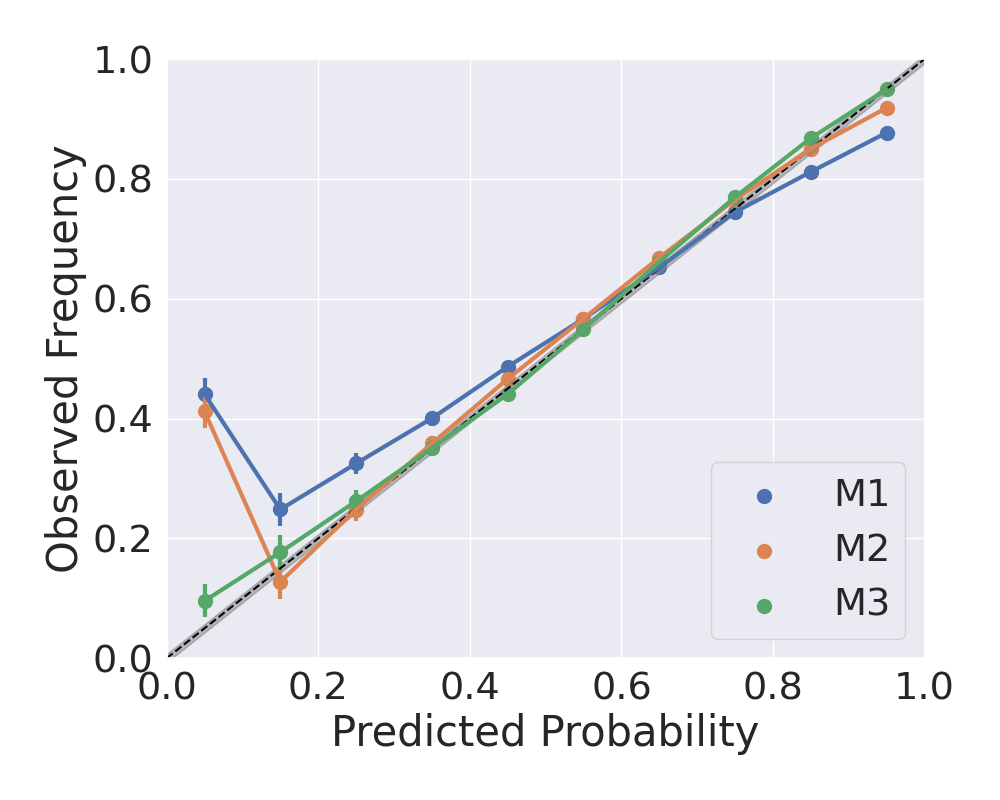} \\
 (a) {Household Group 1.} & (b) {Household Group 3.}\\
\end{tabular}}
\caption{Frequency calibration summaries for 3 DLMMs predicting \$ spend at the category level, averaged across the 2{,}000 households within each household group.}
\label{fig:GRP2-calib}
\end{figure}

Across household groups and metrics, M3 with the simultaneous predictor outperforms models with lagged predictors.  The ZAPE metric, in particular, explicitly evaluates how well the DLMM predicts zeros, or the occurrence of no-spend in the category. M3 also improves forecast calibration across all households and weeks, relative to the models with the lagged predictors (\autoref{fig:GRP2-calib}).  The modeling decomposition generally improves calibration at this modeling level, as we have already conditioned out zeros that are a result of no return to purchase at all.  While calibration of the new DLMM is quite good across all models and all weeks (\autoref{fig:GRP2-calib}),  predicting the occurrence of zeros for a specific week can still be challenging, and thus  conditioning out zeros representing no return at higher levels in the modeling hierarchy can improve   forecasts at lower levels.

\begin{table}[h]
\caption{Forecast summaries for 3 DLMMs predicting \$ spend at the category level, giving median (and 25, 75th percentiles) values across the 2{,}000 households within each household group.}
\vspace{-.5cm}
\begin{center}
\scalebox{0.9}{\begin{tabular}{c|l|c|c|c}
HH Group & Metric & M1: Lagged Cat. & M2: Lagged Global & M3: Simultaneous Global \\\hline
\multirow{ 3}{*}{1} & MAD & 4.31, (2.83, 6.03)  & 4.21, (2.76, 5.83)  & $\bm{3.39}$, (2.42, 4.37) \\
& MAPE & 0.51, (0.45, 0.58) & 0.51, (0.44, 0.57)  & $\bm{0.42}$, (0.37, 0.47)  \\
& ZAPE  & 0.61, (0.54, 0.69) & 0.60, (0.54, 0.68) & $\bm{0.48}$, (0.43, 0.54) \\\hline
\multirow{ 3}{*}{2} & MAD &  4.14, (3.27, 5.35) & 3.47, (2.62, 4.46)  & $\bm{2.82}$, (2.27, 3.39)  \\
& MAPE & 0.53, (0.47, 0.60) & 0.48, (0.44, 0.53) & $\bm{0.41}$, (0.37, 0.45)  \\
& ZAPE &0.68, (0.60, 0.91) & 0.62, (0.56, 0.70) & $\bm{0.49}$, (0.45, 0.53) \\\hline
\multirow{ 3}{*}{3} & MAD & 2.88, (2.14, 3.78)  & 2.77, (2.09, 3.60)  &  $\bm{2.28}$, (1.83, 2.77) \\
& MAPE & 0.46, (0.41, 0.52)  & 0.45, (0.40, 0.50)  & $\bm{0.39}$, (0.35, 0.43)  \\
& ZAPE & 0.65, (0.58, 0.76) & 0.63, (0.57, 0.74) & $\bm{0.49}$, (0.45, 0.53)  \\
\end{tabular}}
\end{center}
\label{tab:grp2-6}
\end{table}%

\FloatBarrier 
\subsubsection{Sub-Category Level}

At the next level in the hierarchy, a DLMM defines p(log Total Spend by Sub-Category $\vert$ Return in Category, Category log Total Spend). Simultaneous predictors again improve forecasting accuracy.  Candidate sub-category models M1, M2, M3 each have a random-walk local level and one model-specific dynamic regression term. Here  
M1 has log total spend at the last return at the sub-category level,  M2 has log total spend at the last return in the category level, while 
M3 has  log total spend for the current week at the category level. As shown in~\autoref{tab:grp1-52},  use of the simultaneous predictor in M3 very substantially improves forecast accuracy at the sub-category level. Calibration (not shown) is also improved by M3. 
\begin{table}[htp]
\caption{Forecast summaries for 3 DLMMs predicting \$ spend at the sub-category level, with details as in~\autoref{tab:grp2-6}.}
\begin{center}
\vspace{-0.5cm}\scalebox{0.9}{
\begin{tabular}{c|l|c|c|c}
HH Group & Metric & M1: Lagged Sub-Cat. & M2: Lagged Cat. & M3: Simultaneous Cat. \\\hline
\multirow{ 3}{*}{1} & MAD & 3.69, (2.52, 5.36)  & 3.67, (2.51, 5.36)  & $\bm{2.15}$, (1.63, 2.72) \\
& MAPE & 0.53, (0.43, 0.63) & 0.52, (0.43, 0.63)  & $\bm{0.32}$, (0.27, 0.37)  \\
& ZAPE  & 0.67, (0.56, 1.00) & 0.67, (0.56, 0.99) & $\bm{0.39}$, (0.33, 0.46) \\\hline
\multirow{ 3}{*}{2} & MAD &  2.64, (1.97, 3.48) & 2.61, (1.98, 3.43)  & $\bm{1.88}$, (1.53, 2.31)  \\
& MAPE & 0.43, (0.37, 0.50) & 0.43, (0.37, 0.49) & $\bm{0.31}$, (0.26, 0.35)  \\
& ZAPE & 0.63, (0.54, 0.80) & 0.63, (0.54, 0.80) & $\bm{0.39}$, (0.33, 0.45) \\\hline
\multirow{ 3}{*}{3} & MAD & 2.81, (2.18, 3.65)  & 2.80, (2.18, 3.64)  &  $\bm{1.74}$, (1.37, 2.17) \\
& MAPE & 0.49, (0.42, 0.58)  & 0.49, (0.41, 0.58)  & $\bm{0.29}$, (0.24, 0.34)  \\
& ZAPE & 0.62, (0.51, 1.94) & 0.62, (0.51, 1.96) & $\bm{0.39}$, (0.32, 0.46)  \\
\end{tabular}}
\end{center}
\label{tab:grp1-52}
\end{table}%

\subsubsection{Item Level}

At the finest level of individual household-item pairs,  a DCMM defines p(Item Quantity $\vert$ Return Sub-Category, Sub-Category log Total Spend, Discount Percent). Again we find that 
simultaneous predictors improve forecast accuracy.   The 3 models evaluated extend the comparisons made at higher levels, each having a trend term as well as the predictors log total spend and discount percent.  M1 includes lagged predictors at the sub-category level, M2 includes lagged predictors at the item level, while M3 considers simultaneous predictors at the sub-category level. Forecast accuracy metrics show the comparisons in \autoref{tab:item-62-simul}, where the improvements in M3 are very substantial.  Again, similar findings emerge for calibration at the item level. 
\begin{table}[htp]
\caption{Forecast summaries for 3 DCMMs predicting item quantity at the final level, with details as in~\autoref{tab:grp2-6}.}
\begin{center}
\vspace{-0.5cm}\scalebox{0.9}{
\begin{tabular}{c|l|c|c|c}
HH Group & Metric & M1: Lagged Sub-Cat. & M2: Lagged Item & M3: Simultaneous Sub-Cat. \\\hline
\multirow{ 3}{*}{1} & MAD & 1.46, (1.00, 2.01)  & 1.46, (0.99, 2.01)  & $\bm{1.00}$, (0.64, 1.40) \\
& MAPE & 0.54, (0.38, 0.71) & 0.53, (0.39, 0.71)  & $\bm{0.35}$, (0.25, 0.44)  \\
& ZAPE  & 0.55, (0.44, 0.68) & 0.54, (0.43, 0.67) & $\bm{0.40}$, (0.31, 0.48) \\\hline
\multirow{ 3}{*}{2} & MAD &  1.14, (0.83, 1.52) & 1.12, (0.82, 1.51)  & $\bm{0.87}$, (0.60, 1.17)  \\
& MAPE & 0.45, (0.30, 0.61) & 0.45, (0.30, 0.60) & $\bm{0.31}$, (0.23, 0.39)  \\
& ZAPE & 0.48, (0.38, 0.59) & 0.48, (0.38, 0.58) & $\bm{0.37}$, (0.30, 0.45) \\\hline
\multirow{ 3}{*}{3} & MAD & 1.11, (0.80, 1.49)  & 1.11, (0.80, 1.49)  &  $\bm{0.78}$, (0.53, 1.09) \\
& MAPE & 0.48, (0.31, 0.66)  & 0.48, (0.31, 0.65)  & $\bm{0.28}$, (0.19, 0.36)  \\
& ZAPE & 0.50, (0.40, 0.61) & 0.50, (0.40, 0.61) & $\bm{0.35}$, (0.27, 0.43)  \\
\end{tabular}}
\end{center}
\label{tab:item-62-simul}
\end{table}%

\subsection{Modeling Decomposition}

In addition to enabling sharing of  simultaneous predictors across levels, the modeling decomposition induces a partial tree-like structure related to zero sales events, and this yields computational and statistical efficiency.  If a specific household does not return in a given week,  the problem of  predicting sales of any item is obviated.  Further, there can be zeros for multiple  weeks for specific items as noted in Section~\ref{sec:data}; then the interest is in  forecasting when there will be a non-zero sale, how many items will be purchased then, and how price-sensitive the household is.  The modeling decomposition facilitates these goals by utilizing the structure of the data to condition out zeros at lower levels of the model, and this ultimately improves forecast accuracy.   

Examples are given for the low spending Household Group 3 in which the proportion of zero purchases is highest.  We compare the full analysis  with 
that based on non-hierarchical, individual household-item level models, referred to as \lq\lq direct'' models.  For fair comparisons, all models  use only lagged predictors. 
Two direct models M1 and M2   are DCMMs for   p(Item Quantity); they each include a local intercept but use different lagged predictors of log total spend and discount percent: M1 uses these at the sub-category level, and M2 at the item level.    Corresponding  hierarchical models, M3 and M4, define p(Item Quantity $\vert$ Return Sub-Category, Sub-Category log Total Spend, Discount Percent) with intercept terms and  lagged predictors as in M1 and M2, respectively. 
\autoref{tab:direct} shows typical results for 3 of the items, indicative of the benefits of the hierarchical structure in outperforming direct models. 
\begin{table}[htp]
\caption{Forecast summaries for the households in Group 3 using DCMMs at the item level. Hierarchical models M3 and M4 exploit the modeling decomposition, direct models M1 and M2 do not. Format as in earlier tables.}
\begin{center}
\vspace{-0.5cm}\scalebox{0.9}{
\begin{tabular}{c|c|c|c|c}
Item & M1: Direct Sub-Cat. & M2: Direct Item & M3: \text{Decomp.} Sub-Cat. & M4: \text{Decomp.} Item \\\hline
A & 0.60, (0.47, 0.75) & 0.61, (0.47, 0.75) & $\bm{0.50}$, (0.40, 0.61) & $\bm{0.50}$, (0.40, 0.61) \\
C & 0.62, (0.41, 0.81) & 0.62, (0.40, 0.81) & $\bm{0.37}$, (0.24, 0.50) & $\bm{0.38}$, (0.25, 0.52) \\
E &  0.87, (0.59, 1.00) & 0.85, (0.58, 1.00) & $\bm{0.52}$, (0.39, 0.70) & $\bm{0.53}$, (0.42, 0.67)\\
\end{tabular}}
\end{center}
\label{tab:direct}
\end{table}%

 \vspace{-0.8cm}
\subsection{Multi-Scale Modeling with Price Discount Predictors}\label{subsec:multi}

We now turn to models that integrate item-specific price discount information at household and household group levels, 
extending the hierarchical modeling framework to share information across households in predicting spends on specific items.
This defines a multi-scale approach that shares information across items {\em and} households via price discount information at the item level. 
Specific predictors based on price discounts are constructed as follows. The {\em potential  discount} amount is (i) the observed discount amount for each week the item is purchased, or (ii)  an imputed value of discount on offer if an item was not purchased by the household that  week.  This predictor is available in external simultaneous form, as discounts due to specific promotions are set weeks in advance. 
We compare two DCMMs, M1 and M2, for  p(Item Quantity $\vert$ Return Sub-Category, Sub-Category log Total Spend); 
each has a local intercept, a predictor given by the simultaneous log total spend at the sub-category level, and a predictor based on  potential discount percentage.  The discount information in M1 is specific to the item and household, while the discount information in M2 is the average   across all households in the   group.  Hence M2 shares information across items via the simultaneous log total spend at the sub-category level predictor and across households via the aggregate discount percentage predictor.  The summary comparisons in~\autoref{tab:item-multi-scale} for item A are typical of the results.  For the majority of households, including multi-scale discount information (M2) improves forecast accuracy across metrics; this more general multi-scale view of sharing information across data ``dimensions'' can lead to improvements in forecasting at the individual level.  These results are typical and bear out the utility of the approach, and suggest that additional extensions could consider sharing multi-scale information across other dimensions, including across time, in future extensions. 
\begin{table}[htp]
\caption{Forecast summaries using hierarchical, item-level DCMMs to predict sales of item A,  comparing price discount predictors in two models:  M1 has household-specific discount percent while M2 has a multi-scale aggregate discount percent predictor. Format as in earlier tables.}
\begin{center}
\vspace{-0.5cm}\scalebox{0.9}{
\begin{tabular}{c|l|c|c} 
HH Group & Metric & M1: HH Discount & M2: Multi-Scale Discount \\\hline
\multirow{ 3}{*}{1} & MAD & 1.06, (0.71, 1.49) & $\bm{0.97}$, (0.66, 1.30) \\ 
& MAPE & 0.39, (0.26, 0.50)  & $\bm{0.35}$, (0.25, 0.42)  \\ 
& ZAPE  & 0.42, (0.33, 0.51) & $\bm{0.38}$, (0.31, 0.44)   \\\hline 
\multirow{ 3}{*}{2} & MAD & 0.91, (0.65, 1.19) &  $\bm{0.85}$, (0.61, 1.10)    \\ 
& MAPE & 0.34, (0.23, 0.43) & $\bm{0.31}$, (0.22, 0.38)    \\
& ZAPE & 0.39, (0.31, 0.46) & $\bm{0.37}$, (0.30, 0.42)  \\\hline
\multirow{ 3}{*}{3} & MAD &  0.82, (0.57, 1.11) & $\bm{0.78}$, (0.55, 1.05) \\
& MAPE & 0.30, (0.19, 0.40) & $\bm{0.28}$, (0.19, 0.37)   \\
& ZAPE & 0.36, (0.28, 0.44) & $\bm{0.35}$, (0.28, 0.41)  \\
\end{tabular}}
\end{center}
\label{tab:item-multi-scale}
\end{table}%

\vspace{-0.8cm}

\subsection{Household-Item Price Sensitivity}\label{subsec:individ}

The discussion so far has demonstrated aspects of forecast improvements based on the modeling decomposition, with results aggregated across households. We now turn to the key question of item-specific price sensitivity for each individual household.   Here we compare models with and without discount information as a dynamic predictor. Households showing improved forecast accuracy with inclusion of discount information will be identified as price-sensitive.  Resulting inferences will feed into  decision processes for individualized discount offers.  Further, as price sensitivity is modeled dynamically, we are able to monitor 
if and how the sensitivities change over time.

\begin{figure}
\centering
\begin{subfigure}{.4\textwidth}
  \centering
  \includegraphics[width=\linewidth]{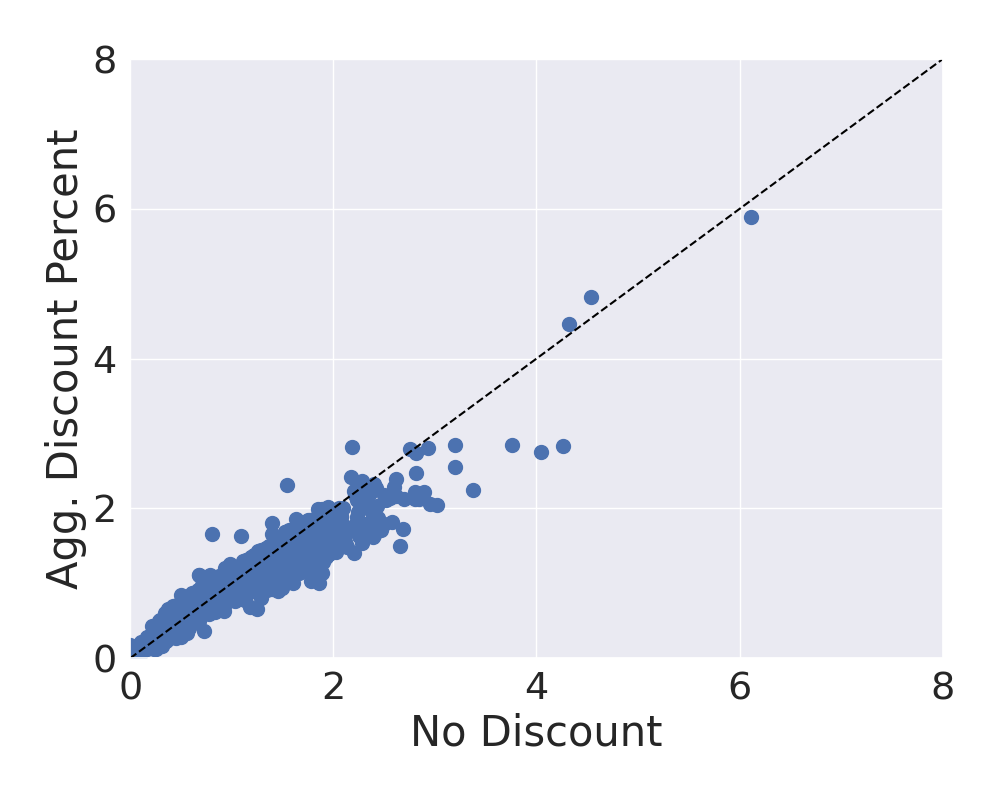}
  \caption{MAD.}
 \label{fig:1b}
\end{subfigure}%
\begin{subfigure}{.4\textwidth}
  \centering
  \includegraphics[width=\linewidth]{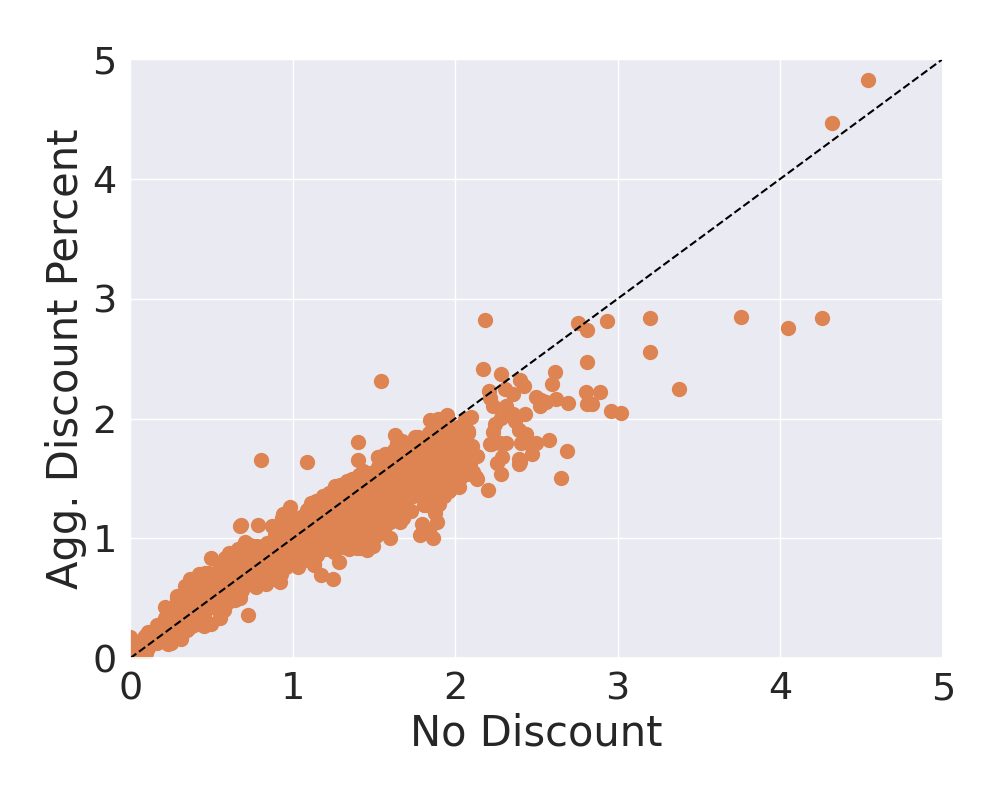}
  \caption{ZAPE.}
 \label{fig:3b}
\end{subfigure}
\caption{MAD and ZAPE metrics for individual households in Household Group 1 from DCMMs  with and without a discount predictor for Item A.  Points represent households;  those
below the diagonal have forecasts that are improved by including discount information.}
\label{fig:scatter-price}
\end{figure}

Summaries come from comparison of two item-level DCMMs for p(Item Quantity $\vert$ Return Sub-Category, Sub-Category log Total Spend).  Both models include a local level and a simultaneous predictor for log total spend at the sub-category level.  The models differ only in that one includes   the aggregate discount percentage across households, found to be a useful predictor in Section~\ref{subsec:multi}, while the other does not.  Household-specific MAD and ZAPE measures are shown in \autoref{fig:scatter-price}.  Each point represents an individual household. For households below the diagonal, including discount information improves forecast accuracy and are households that have the potential to be price-sensitive so are perhaps candidates for more customized promotions. 

\begin{figure}[htbp!]
\centering
\includegraphics[width=.7\linewidth]{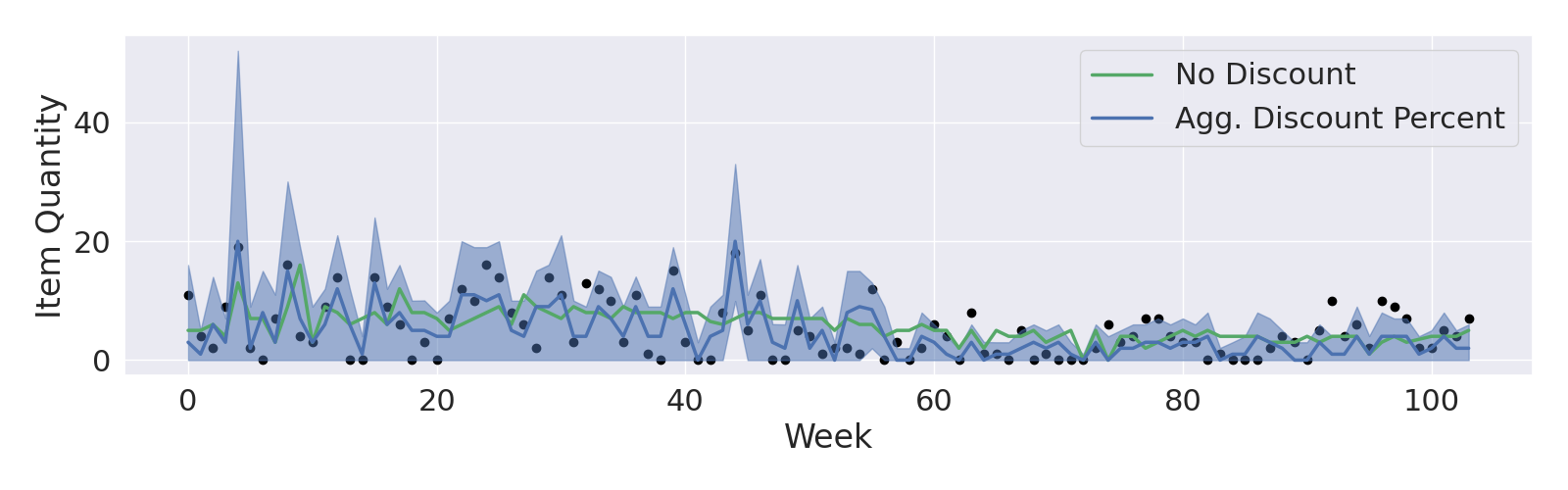} \\
(a) {1-Step Ahead Forecasts.} \\
\includegraphics[width=.7\linewidth]{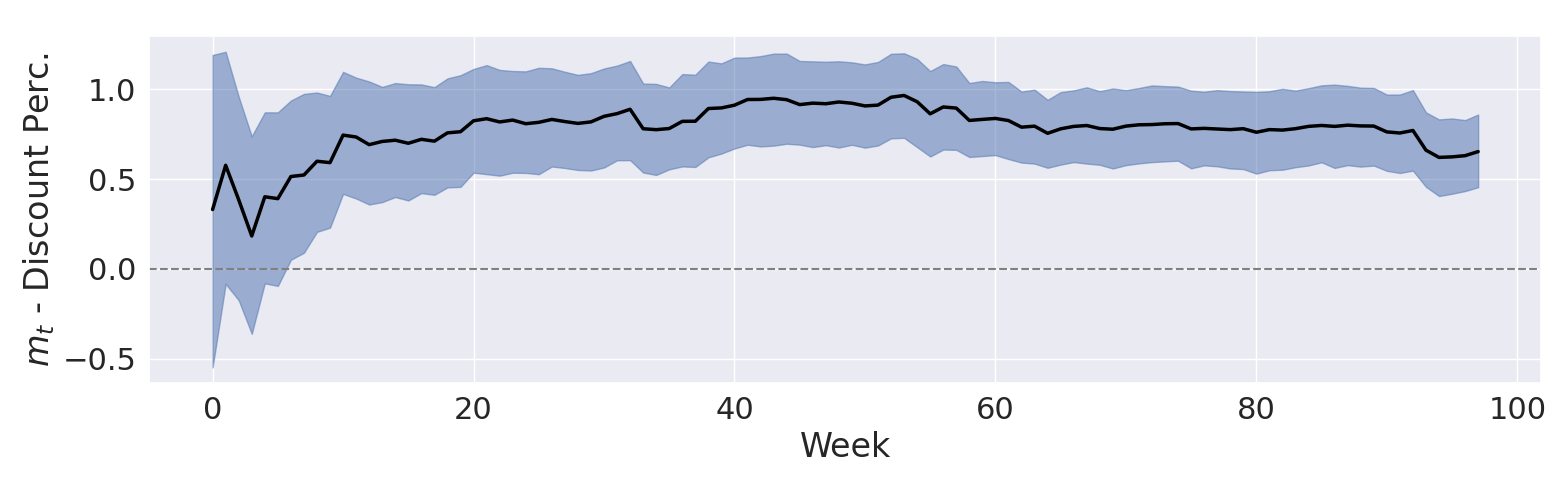} \\
(b) {Discount Sensitivity.} \\
\caption{(a) MAD optimal point forecasts and  90\% prediction intervals for  Item A purchases of one price-sensitive household, with and without discount information; (b) on-line posterior mean and  90\% intervals for the  state vector element  corresponding to the discount predictor.}\label{fig:HH1-discount}
\end{figure}

\begin{figure}[h!]
\centering
\includegraphics[width=.7\linewidth]{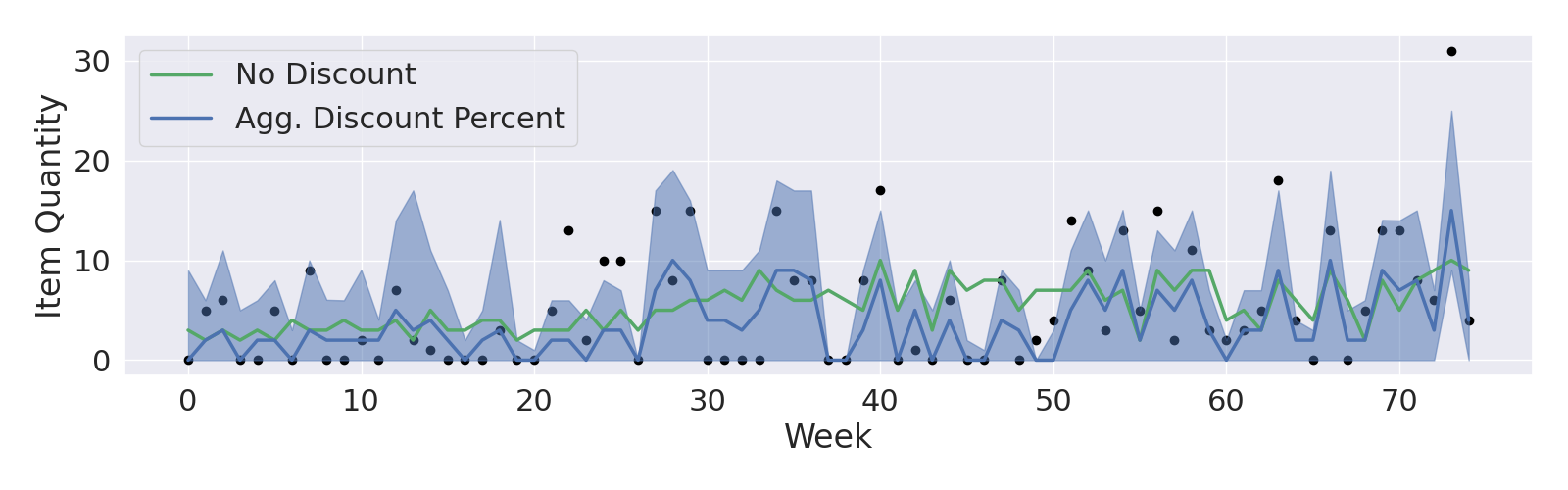}\\
(a) {1-Step Ahead Forecasts.} \\ 
\includegraphics[width=.7\linewidth]{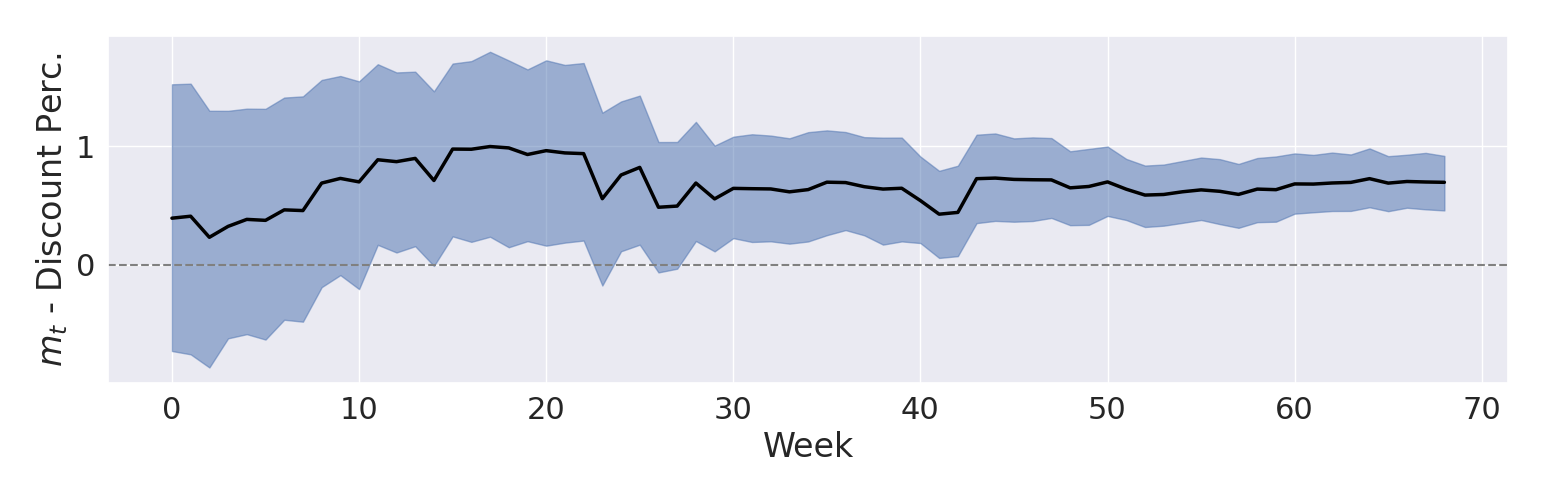} \\
(b) {Discount Sensitivity.} \\
\caption{ZAPE optimal point forecasts and other summaries for second price-sensitive household, with format as in~\autoref{fig:HH1-discount}.}
\label{fig:HH2-discount}
\end{figure}

Figures~\ref{fig:HH1-discount} and~\ref{fig:HH2-discount} show summary results for Item A purchases of 
 two price-sensitive, high spending households from Group 1. The visual presentation shows improvements with the inclusion of discount information. For each household,  the state vector coefficient on discount percentage-- the discount sensitivity in these models-- is inferred to be positive with high probability across the time period,  indicating that higher discount percentages are associated with increased items purchased for these households. These are therefore households that may respond positively to  additional discount offers.

\FloatBarrier

\begin{figure}[htbp!]
\centering
\includegraphics[width=.7\linewidth]{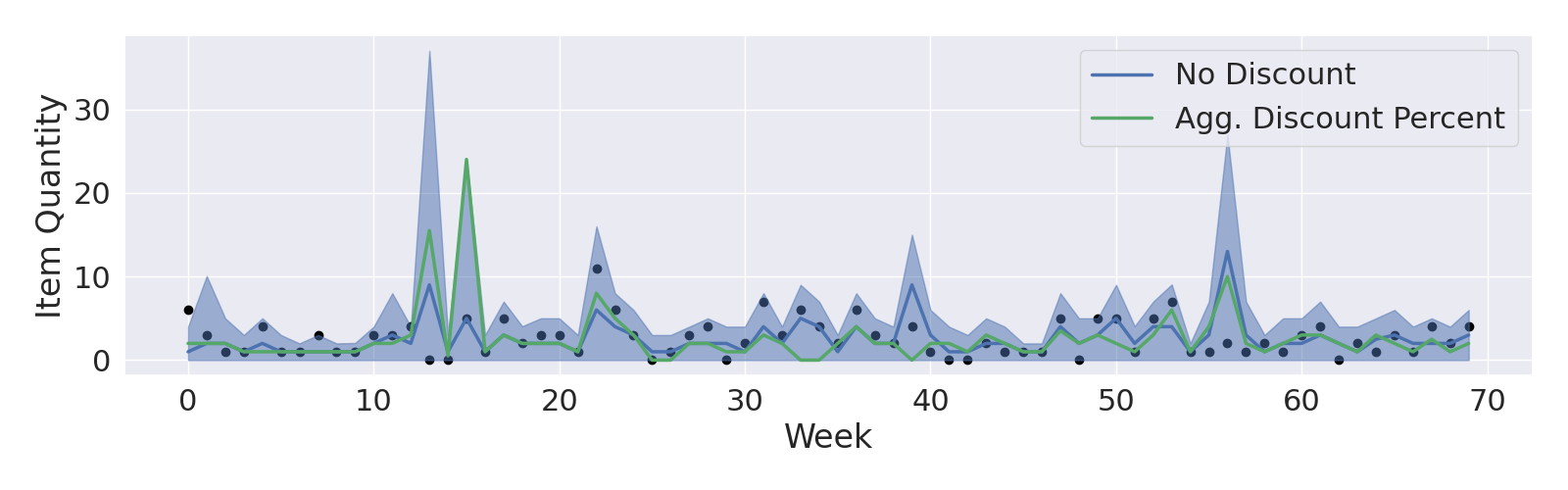}\\
(a) {1-Step Ahead Forecasts.} \\ 
   \includegraphics[width=.7\linewidth]{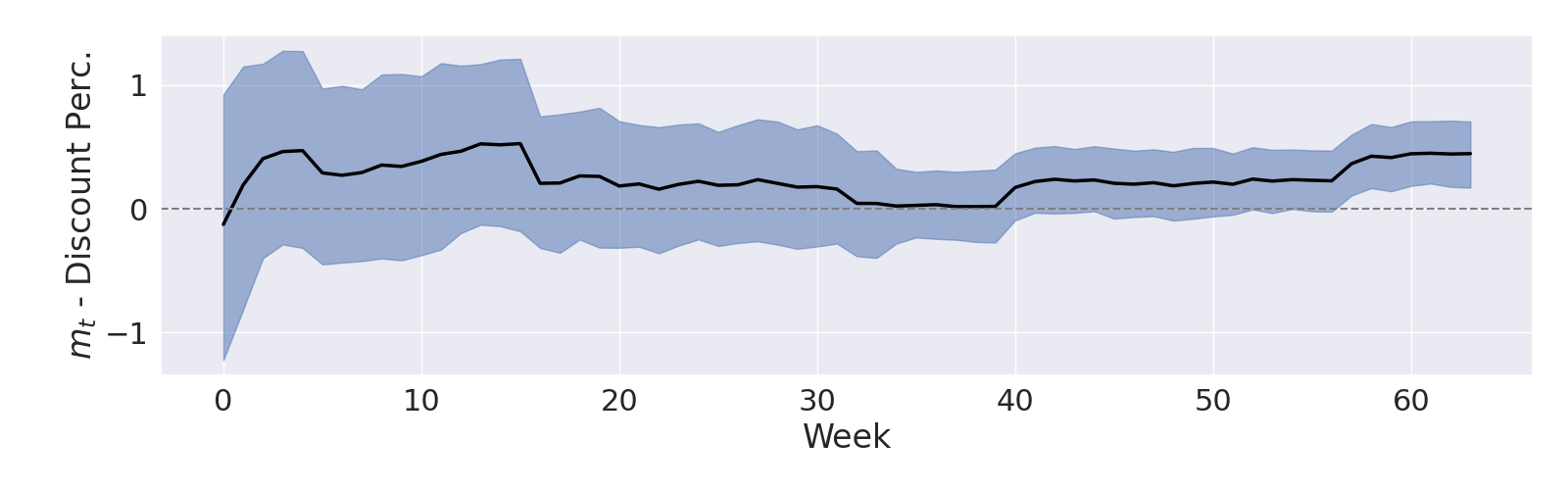} \\
(b) {Discount Sensitivity.} \\
\caption{(a) ZAPE optimal point forecasts and other summaries for a third household that is not  price-sensitive, with format as in~\autoref{fig:HH1-discount}.}\label{fig:HH3-discount}
\end{figure}

A third Group 1 household that is not price-sensitive provides contrast; see~\autoref{fig:HH3-discount}.  Item A forecasts for the model without discount information are more accurate than the corresponding forecasts for the model with discount information; correspondingly,  the discount sensitivity state element is inferred as insignificant over time.  This  household is unlikely to be responsive to discount offers in the normal ranges.  

The results across all households and details for each item-household pair show the use of models with simultaneous predictors; they generally improve predictions while identifying individual households that are price-sensitive for specific items, estimating how this price sensitivity changes over time.  This opens the door to direct discount interventions for price-sensitive households, enabled  by the use of interpretable models at each modeling level.   The framework also facilitates  decision-theoretic approaches to selection of optimal discount percent to offer to each price-sensitive household,  and to update the strategy over time.

\subsection{Some Forecast Assessments in Simultaneous Modeling\label{subsec:simul}}

The analysis at lower levels of the hierarchy  involves  simultaneous predictors from higher levels. Forecast information on these predictors cascades down from the global to finer levels, defining the coupling of sets of models in the hierarchy.  This can be done using full ensembles of synthetic values generated from higher level predictive distributions, and/or using point forecasts.  We detail this further assuming the simultaneous forecasts are in terms of medians or means.  For one household-item pair in one week, predicting one week ahead begins with the Bernoulli DGLM for p(Return) at the global level. This defines a  forecast that is propagated to serve as a predictor in  the DLMM for p(Global log Total Spend $\vert$ Return) along with the known predictor of lagged global log total spend.  This DLMM provides a forecast of global spend that, together with the forecast Return, is projected down to the 
category level DLMM of p(Category log Total Spend $\vert$ Return, Global log Total Spend).  The process continues with forecast values projected to the 
sub-category level DLMM of p(Sub-Category log Total Spend $\vert$ Return Category, Category log Total Spend).  Then, finally, the forecast from the sub-category model is projected to define the DCMM of p(Item Quantity $\vert$ Sub-Category Return, Sub-Category log Total Spend).   

\begin{figure}
\centering
\begin{tabular}{cc} 
 \includegraphics[width=.4\linewidth]{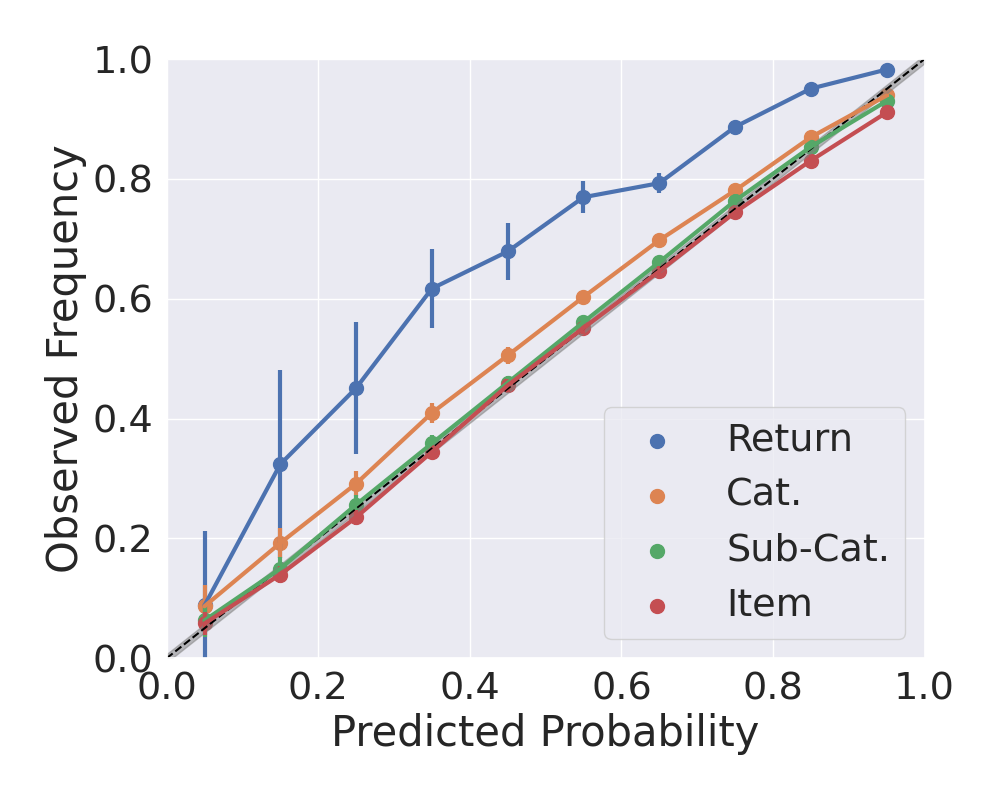} &
 \includegraphics[width=.4\linewidth]{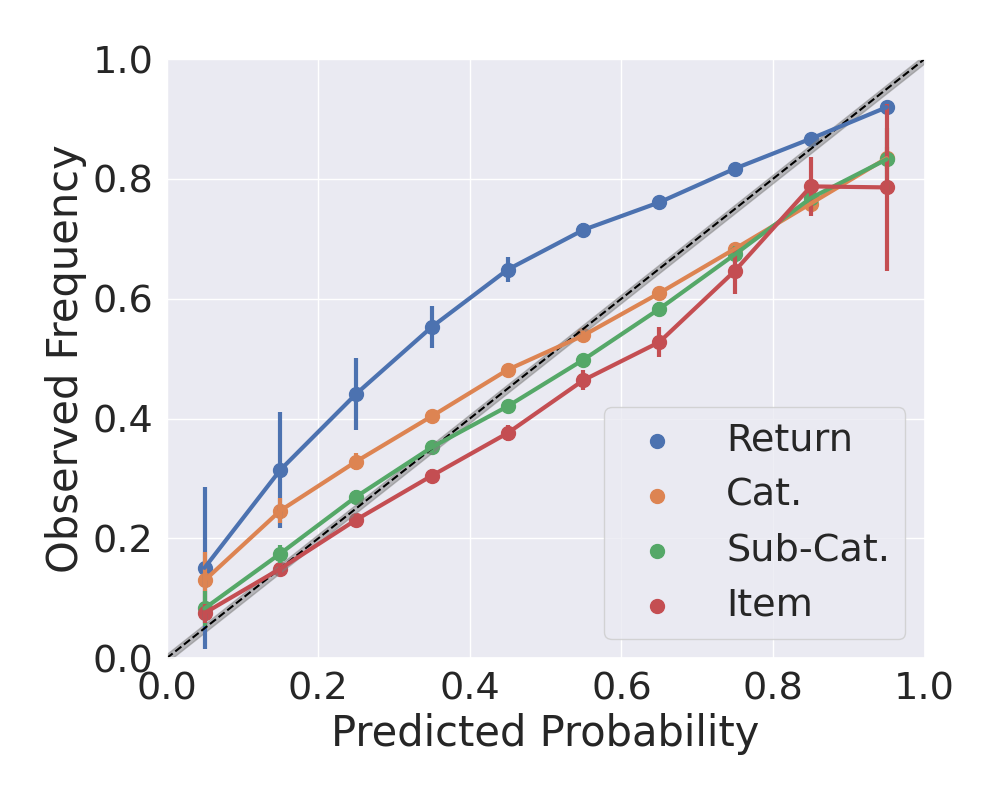} \\
 (a) Household Group 1. & (b) Household Group 3.\\
 \end{tabular}
\caption{Frequency calibration within each of the model levels for the high-spending (Group 1) and low-spending (Group 3) household groups  for Item A, using forecast means for projected simultaneous predictors.}
\label{fig:calib-simul}
\end{figure}

Some summary evaluations are  presented for Item A by household group. 
We focus on aspects of prediction accuracy for one of the most challenging components of the analysis, that of predicting 
household return at each of the the global, category, sub-category and item levels. Calibration plots (\autoref{fig:calib-simul}) and confusion matrices (\autoref{tab:simul-confusion}) speak to this. Across all households and weeks, predicting global return remains the key challenge, while calibration at subsequent levels is generally very good. The confusion matrices also show that overall, the modeling decomposition tends to do a good job at predicting return for each household.  However, at lower levels in the modeling decomposition, where more predictors are propagated via  forecasts, projecting  forecast means tends to over-predict return, while projecting  forecast medians tends to under-predict return~(\autoref{tab:simul-confusion}).  Nevertheless, the overall forecast accuracy using projected forecasts 
for simultaneous predictors compares very well with that based on the unobtainable \lq\lq gold-standard''-- assuming the simultaneous predictors are known (at their future realized values); see \autoref{tab:simul-forecast}. This is a strong testament to the utility of the hierarchical model decomposition and the overall approach.

\begin{table}
\caption{Confusion matrices in predicting return at each modeling level for household Group 1 and Item A, using forecast means or medians for projected simultaneous predictors.  Entries are the proportions of households and weeks which fall into each cell in the confusion matrix; $f_t$ denotes the point forecast  and $y_t$ the observed value; $f_t = 0$ or $y_t = 0$ indicates no return (forecast or actual). 
}
\begin{center}
\vspace{-0.5cm}\scalebox{0.9}{
\begin{tabular}{cc|cc|cc|cc}
 & &  \multicolumn{2}{c|}{Return} & \multicolumn{2}{c|}{Category} & \multicolumn{2}{c}{Sub-Category} \\
  & & $f_t > 0$ & $f_t = 0$ & $f_t > 0$ & $f_t = 0$ & $f_t > 0$ & $f_t = 0$ \\\hline
 \multirow{ 2}{*}{Mean} & $y_t > 0$ & 0.83 & 0.13 & 0.58 & 0.14 & 0.43 & 0.16 \\
 & $y_t = 0$ & 0.03 & 0.01 & 0.21 & 0.07 & 0.29 & 0.13\\\hline
 \multirow{ 2}{*}{Median} & $y_t > 0$ & 0.83 & 0.13 & 0.52 & 0.20 & 0.34 & 0.25 \\
 & $y_t = 0$  & 0.03 & 0.01 & 0.10 & 0.17 & 0.10 & 0.31 \\
\end{tabular}}
\end{center}
\label{tab:simul-confusion}
\end{table}%

\begin{table}[htp]
\caption{Median (25\%,75\%) forecast accuracy assessments.  M1 and M2 use forecasts of simultaneous predictors-- the forecast mean and median, respectively;  M3 is the hypothetical gold-standard using realized values.}
\begin{center}
\vspace{-0.5cm}\scalebox{0.9}{
\begin{tabular}{c|l|c|c|c}
HH Group & Metric & M1: Simul. Mean & M2: Simul. Median & M3: Known \\\hline
\multirow{ 3}{*}{1} & MAD & 0.90, (0.40, 1.48) & 1.19, (0.79, 1.72)  & 0.99, (0.63, 1.40)  \\
& MAPE & 0.43, (0.28, 0.55) & 0.46, (0.34, 0.57)& 0.35, (0.25, 0.44)  \\
& ZAPE  & 0.44, (0.27, 0.56) & 0.50, (0.40, 0.59) &0.40, (0.31, 0.48)  \\\hline
\multirow{ 3}{*}{2} & MAD &  0.71, (0.39, 1.06) & 0.94, (0.65, 1.24) & 0.87, (0.59, 1.16)  \\
& MAPE & 0.37, (0.25, 0.49) & 0.40, (0.29, 0.50)& 0.31, (0.22, 0.39)  \\
& ZAPE & 0.39, (0.25, 0.50) &0.45, (0.36, 0.53) & 0.37, (0.29, 0.45) \\\hline
\multirow{ 3}{*}{3} & MAD & 0.45, (0.23, 0.73) & 0.74, (0.45, 1.00) & 0.77, (0.53, 1.09)  \\
& MAPE & 0.32, (0.20, 0.45) & 0.35, (0.23, 0.47)   & 0.28, (0.19, 0.36) \\
& ZAPE & 0.29, (0.16, 0.42) & 0.40, (0.30, 0.48) & 0.35, (0.26, 0.43)  \\
\end{tabular}}
\end{center}
\label{tab:simul-forecast}
\end{table}%

 \FloatBarrier

\section{Summary Comments\label{sec:conc}}

We have developed and summarized a case study in Bayesian modeling and forecasting of large, heterogeneous,  individual-level time series data using a decouple/recouple strategy. The applied setting  and extended case study is defined by central and topical issues of consumer behavior modeling and sales forecasting in a retail context, with intimate links to the day-to-day issues of supply chain management and interests in personalized (household-customer level) prediction for decisions. 
%

The applied advances are based on customized, multi-scale models that overlay the inherent hierarchical  nature of the context: consumers decide to visit a store, when they are there they spend, and spending outcomes cascade through broad categories of goods, to refined sub-categories, and ultimately to specific items on sale. How much an individual household spends overall is a useful predictor of potential outcomes-- spend and numbers of items at the finest level-- so that cascading information down the hierarchy is key to decoupling levels.  The recoupling is then inherently defined through simultaneous outcomes; the amount spent at one level  defines a predictor for finer levels. 
We have presented selected summaries of  results that particularly highlight the role and relevance of the model decomposition reflecting these simultaneous predictors and the ability to increase forecast accuracy.   We have also exemplified the use of the model in identifying individual households that are price-sensitive, opening the path to integrating these interpretable probabilistic models with decision  analysis focused on (dynamic) selection of personalized pricing/discounts at the individual item level.  Importantly, we find that our positive results generalize across very heterogenous groups of households and items.

In terms of empirical forecast accuracy, a main challenge is that of predicting return or no return at each modeling level.  Extended or alternative models   could be considered for these binary prediction  components.  There is also opportunity for additional multi-scale extensions. We might, for example, consider aggregation of household purchases over time, or over households at multiple modeling levels that involve linkages based on household demographics or past behavior.  

The selective analysis of forecast accuracy in  simultaneous modeling in Section~\ref{subsec:simul} is based on projecting point forecasts of simultaneous predictors from higher to lower levels of the hierarchy.  We have noted that this is trivially extended to 
project  full forecast ensembles-- samples from predictive distributions at each level-- that will, of course, more formally represent the uncertainties as they propagate (and typically increase) down the hierarchy.  While not shown here, this underlies the holistic Bayesian approach with the resulting ability to average over the uncertain simultaneous predictors (extending and building on~\citealp{BerryWest2018DCMM} and~\citealp{BerryWest2018DBCM}, for example). 
  This may in some cases have practical impact on point forecast accuracy and calibration at the finest level, though the impact is context-specific and to be explored case-by-case in empirical studies. 
 One main concern in generating large ensembles is, of course, the implied computational burden; when
the  full multi-scale model is to be run at each time point with large Monte Carlo ensembles of simultaneous predictors at each level, this quickly becomes a challenge.  On this theme, some recent developments in alternative, partially analytic, approaches to cascading forecast uncertainties in multi-scale frameworks
~\citep{LavineCronWest2020factorDGLMs} will be worth exploring for adaptation to the current setting.   



\section*{Acknowledgements}
 
The research reported here was partly supported by $84.51^\circ$. Our research has benefited from discussions with  $84.51^\circ$ Research Scientist
Christoph Hellmayr.  Any opinions, findings and conclusions or recommendations expressed in this paper do 
 not necessarily reflect the views of $84.51^\circ$.



\bibliographystyle{plainnat} 
\bibliography{HHmodelsRefs}
 
 \break\newpage
\appendix

\section{Sequential Learning and Forecasting}

\subsection{DGLMs: Dynamic Generalized Linear Models}

Sequential learning for the DGLM \citep{West-Harrison} proceeds as follows for the time $t-1$ evolve-predict-update cycle (following \citet{Berry:2019}):
\begin{enumerate}
\itemsep=0pt
\item Posterior at $t-1$: $\left(\bm \theta_{t-1} \vert \mathcal D_{t-1}, \mathcal I_{t-1}\right) \sim \left(\bm m_{t-1}, \bm C_{t-1}\right).$
\item Prior at $t$: $\left(\bm \theta_{t} \vert \mathcal D_{t-1}, \mathcal I_{t-1}\right) \sim \left(\bm a_t, \bm R_t\right)$ with $\bm a_t = \bm G_t\bm m_{t-1}$ and $\bm R_t = \bm G_t\bm C_{t-1} \bm G_t' + \bm W_t$.
\item Variational Bayes: $\left(\eta_t \vert \mathcal D_{t-1}, \mathcal I_{t-1}\right)\sim\mbox{CP}\left(\alpha_t, \beta_t\right)$, 
$$p\left(\eta_t \vert \mathcal D_{t-1}, \mathcal I_{t-1}\right) = c(\alpha_t, \beta_t)\mbox{exp}\{\alpha_t\eta_t - \beta_t a(\eta_t)\}.$$ ($c(\cdot, \cdot)$ known function of hyperparameters, depends on exponential family form).
\item Evaluate hyper-parameters $\alpha_t$ and $\beta_t$ such that:
$$\mathbb{E}\left(\lambda_t \vert \mathcal D_{t-1}, \mathcal I_{t-1}\right) = f_t = \bm F_t'\bm a_t\enspace\mbox{and}\enspace\mathbb{V}\left(\lambda_t \vert\mathcal D_{t-1}, \mathcal I_{t-1}\right) = q_t = \bm F_t'\bm R_t\bm F_t.$$
\item Forecast $y_t$ 1-step ahead: $p(y_t \vert \mathcal D_{t-1}, \mathcal I_{t-1}) = b(y_t, \phi)c(\alpha_t, \beta_t)/c(\alpha_t + \phi y_t, \beta_t + \phi)$.
\item Posterior for $\eta_t$: $\left(\eta_t \vert \mathcal D_t\right)\sim \mbox{CP}\left(\alpha_t + \phi y_t, \beta_t + \phi\right)$.
\item Map back to the linear predictor $\lambda_t = g(\eta_t)$: posterior mean $g_t = \mathbb{E}(\lambda_t \vert \mathcal D_t)$ and variance $p_t = \mathbb{V}(\lambda_t \vert \mathcal D_t)$.
\item Posterior at time $t$: $\left(\bm \theta_t | \mathcal D_t\right)\sim\left(\bm m_t, \bm C_t\right)$ given by
$$\bm m_t = \bm a_t + \bm R_t\bm F_t (g_t - f_t)/q_t\enspace\mbox{and}\enspace \bm C_t = \bm R_t - \bm R_t\bm F_t\bm F_t'\bm R_t' (1 - p_t/q_t)/q_t.$$
\end{enumerate}
\noindent This completes the time $t-1$ to $t$ evolve-predict-update cycle.  For all results presented, we specify the following state space priors: $\bm m_0 = \bm 0$ and $\bm C_0 = \bm I$, where $\bm I$ is the identity matrix.

\subsection{Forecasting in Dynamic Count Mixture Models}

The forecast distribution at time $t+k$ for the DCMM is a mixture of the forecast distributions for the independent Bernoulli and shifted Poisson DGLMs. That is, the marginal forecast distributions are
$$p(y_{t+k} | \mathcal{D}_t, \mathcal{I}_t, \pi_{t+k}) = (1 - \pi_{t+k})\delta_0(y_{t+k}) + \pi_{t+k}h_{t, t+k}(y_{t+k}),$$ 

\noindent where $(\pi_{t+k} | \mathcal{D}_t, \mathcal{I}_t)\sim\text{Beta}(\alpha_t^0(k), \beta_t^0(k))$, $\delta_0(y)$ is the Dirac delta function, $h_{t, t+k}(y_{t+k})$ is the density of $y_{t+k} = 1 + x_{t+k}$, where $(x_{t+k} | \mathcal{D}_t, \mathcal{I}_t)\sim\text{NegBinom}(\alpha_t^+(k), \tfrac{\beta^+_0(k)}{1 + \beta_t^+(k)})$, and $\alpha_t^0(k)$ and $\beta_t^0(k)$ are computed from the binary DGLM and  $\alpha_t^+(k)$, $\beta_t^+(k)$ are computed from the shifted Poisson DGLM  \citep{BerryWest2018DCMM}.  We primarily focus on one-step ahead forecasts in the main results, in which case $k = 1$.

\subsection{Forecasting in Dynamic Linear Mixture Models}

In the DLMM, the forecast distribution at time $t+k$ is nearly identical to the forecast distribution in the DCMM, with the exception that $h_{t, t+k}(y_{t+k})$ is now a student-t distribution, when a Beta-Gamma stochastic volatility model is specified for the observation precision, following \citet{West-Harrison}.  


\section{Metric Derivations}

\subsection{Log-T Distributions}

In our modeling context, as often occurs in demand forecasting, we are interested in modeling a log quantity, specifically the log total spend at the global, category and sub-category levels.  We model the log total spend with a DLM, meaning that the predictive distribution is a student-t distribution.  However, we often want to evaluate our models directly on the original dollar scale, rather than the log-scale.  That is, if $y=\log(x)$ and $x\sim T_k(m,v)$, then $y\sim LT_k(m,v)$, which is a heavy-tailed log-T distribution with p.d.f \citep{West2020decisionconstraints}:
\begin{equation*}
p(y) \propto y^{-1}\left(k + \left(\log(y) - m\right)^2/v\right)^{-(k+1)/2}, \enspace y > 0.
\end{equation*}
\noindent This p.d.f decays as an inverse power of $\log(y)$ as $y\rightarrow\infty$ and has pole at zero.  As a result, none of the moments of the log-T distribution exist and expected losses for commonly used losses occurring under the log-T distribution also do not exist \citep{West2020decisionconstraints}.  For example, MAPE and ZAPE loses do not have finite expectations.  In practice, to calculate optimal point forecasts with log-T predictive densities, we can truncate the log-T distribution to bounded values (away from 0 and up to a finite value) to calculate finite expected losses \citep{West2020decisionconstraints}.  

\subsection{Optimal ZAPE Forecasts}

The optimal ZAPE forecasts can be derived following the similar MAPE derivation in \citet{Berry:2019}. Let $y$ be a continuous quantity, where $\pi_0 = \mathbb{P}(y = 0) > 0$.  Additionally, let $p(y)$, and $P(y)$ be the PDF and CDF of $y$, respectively. Let  $g(y) = cy^{-1}p(y)\bm 1(y > 0)$ with CDF $G(y)$ and  $c\geq 1$. Then the optimal forecast under the ZAPE loss function can be derived as follows:
\begin{equation*}\label{mZAPE}
\begin{split}
\mathcal{L}_{ZAPE}(y, f) &= \dfrac{f}{1+f}\times\bm 1(y = 0)+ \bm 1(y > 0)\Bigl| 1 - \tfrac{f}{y}\Bigl|, \\
R(f) &= \int_0^{\infty}\mathcal{L}_{ZAPE}(y, f) p(y)dy  \\
&= \dfrac{f}{1+f}\pi_0 + \int_1^{\infty}|y-f|y^{-1}p(y)dy \\
\dfrac{|y-f|}{y} &= \begin{cases}
  1 - f/y, & \text{if } y\geq f, \\
  f/y - 1, & \text{if } y < f.
\end{cases} \\
\end{split}
\end{equation*}
\begin{equation*}
\begin{split}
R(f) &= \dfrac{f}{1+f}\pi_0 + \int_1^{f}\left(\dfrac{f}{y} - 1\right)p(y)dy + \int_f^{\infty}\left(1 - \dfrac{f}{y}\right)p(y)dy \\
&= \dfrac{f}{1+f}\pi_0 + \int_1^f\dfrac{fc}{yc}p(y)dy - [P(f) - P(1)] + [1 - P(f)] - \int_f^{\infty} \dfrac{fc}{yc}p(y)dy \\
&= \dfrac{f}{1+f}\pi_0 + P(1) + 1 - 2P(f) + \int_1^f \dfrac{f}{c}g(y)dy - \int_f^{\infty}\dfrac{f}{c}g(y) dy \\
&= \dfrac{f}{1+f}\pi_0 + P(1) + 1 - 2P(f) + \dfrac{2f}{c}G(f) - \dfrac{f}{c} \\
\dfrac{\partial R(f)}{\partial f} &= \dfrac{\partial}{\partial f}\left(\dfrac{f}{1+f}\right)\pi_0 - 2p(f) + \dfrac{2}{c}\left(G(f) + cp(f)\right) - \dfrac{1}{c} \\
\dfrac{\partial R(f)}{\partial f} &= \dfrac{\pi_0}{(1+f)^2} + \dfrac{2}{c} G(f) - \dfrac{1}{c}
\end{split}
\end{equation*}

\begin{equation*}
\begin{split}
\dfrac{\partial R(f)}{\partial f} = 0  & \implies  G(f) = \dfrac{1}{2} - \dfrac{\pi_0 c}{2}\left(\dfrac{1}{(1+f)^2}\right) \in [0,1], \\
& \implies \dfrac{1}{2}\left(1 - \dfrac{\pi_0c}{(1+f)^2}\right) \in [0,1] \\
\end{split}
\end{equation*}

Then, the procedure to find the optimal ZAPE forecast, $f^*$, is:

\begin{enumerate}
\item Solve for $\tilde{f}$ via gradient descent:
\begin{equation*}
\begin{split}
\tilde{f} &= f - \alpha \dfrac{\partial R(f)}{\partial f}, \\
\dfrac{\partial R(f)}{\partial f} &= \dfrac{\pi_0}{(1+f)^2} + \dfrac{2}{c} G(f) - \dfrac{1}{c}.
\end{split}
\end{equation*}
\item If $$\dfrac{\pi_0 c}{(1 + \tilde{f})^2} \geq 1,$$
then $f^* = 0$.
\item Otherwise, $f^* = \tilde{f}$.
\end{enumerate}

If $y$ is a non-negative count, we can follow a similar procedure, but where we use a grid search to minimize the risk on values between 0 and the (-1)-median (the median of the distribution $g(y)$ defined above), inclusive.  Thus, for positive counts, we have that the ZAPE optimal forecast is always less than or equal to the MAPE optimal forecast, which is less than or equal to the MAD optimal forecast.


\section{Item Selection Details}

Due to the unavailability of demographic information about households, the following categorization of households is developed, on the basis of the promotion circumstances and buying behaviors, which are defined for every household-item combination.  This categorization is then used to select which items to focus on for modeling.

For every household-item pair: (i,h), i = 1:I, h = 1:H, with I, H being the total number of items being sold and households recorded, define:
\begin{itemize}
\itemsep=0pt
	\item Discount Offered Percentage (DOP): over the span of the 112 weeks recorded, the proportion of weeks when there were promotions offered to household h for item i
	\item Discounted Purchase Percentage (DPP): among the weeks when item i was discounted for household h, the proportion of weeks that household h made a purchase
	\item Regular Purchase Percentage (RPP): among the weeks when item i was at regular price for household h, the proportion of weeks that household h made a purchase
\end{itemize}

These three quantities together define a household space for each item, whose domain is a unit cube centering around the origin, with various sections of the cube corresponding to different purchasing behaviors.  We can then form four household categories based on this overall purchasing behavior for each item, $i$, and possible discount actions to take for each category:

\begin{enumerate}
\itemsep=0pt
	\item Habit and loyalty for item $i$ are established.\\
	Actions: Maintain the relationship and occasionally compensate for their loyalty to item $i$.
	
	\item Promotion sensitivity and interests in item $i$ are detectable---which is the ideal group of customers to model the price sensitivity.\\
	Actions: It is interesting to find the amount of promotions of item $i$ that generates the most profits, which depends on the distribution of sales and the quantity being optimized. 
	
	\item Promotions are not available.\\
	Actions: Explore and experiment with these customers by delivering promotions of item $i$.
	
	\item Lack of interest in item $i$ or disregard for the promotions is noticeable.\\
	Actions: Check the validity of the promotions sent out. If they are disregarded, stop the promotions of item $i$.
\end{enumerate}

After this categorization, we select items based on which items have a large proportion of households in category 2 above, indicating that there are many households for these specific items that are price sensitive.  Additional summary statistics about each item selected are given in \autoref{tab:item-info}.

\begin{table}[htp]
\caption{Sub-Category and Category for each item modeled.  Also, the number of households (out of 2000 for each household group) in each household group which return to purchase each item more than 10 weeks out of the 112 weeks; these are the households that are modeled for each item.}
\begin{center}
\vspace{-0.5cm}\scalebox{0.9}{
\begin{tabular}{c|c|c|c|c|c}
Item & Sub-Category & Category & \# Group 1 & \# Group 2  & \# Group 3 \\\hline
A & 2c & 2 & 1806 & 1877 & 1808 \\
B & 2c & 2 & 1724 & 1795 & 1709 \\
C & 2b & 2 & 1029 & 865 & 511 \\
D & 2a & 2 & 710 & 470 & 296 \\
E & 3a & 3 & 1033 & 851 & 521 \\
F & 1a & 1 & 342 & 268 & 101 \\
\end{tabular}}
\end{center}
\label{tab:item-info}
\end{table}%

\section{Additional Results}

\subsection{Global Return}

\begin{figure}[h]
\centering
\begin{subfigure}{.30\textwidth}
  \centering
  \includegraphics[width=\linewidth]{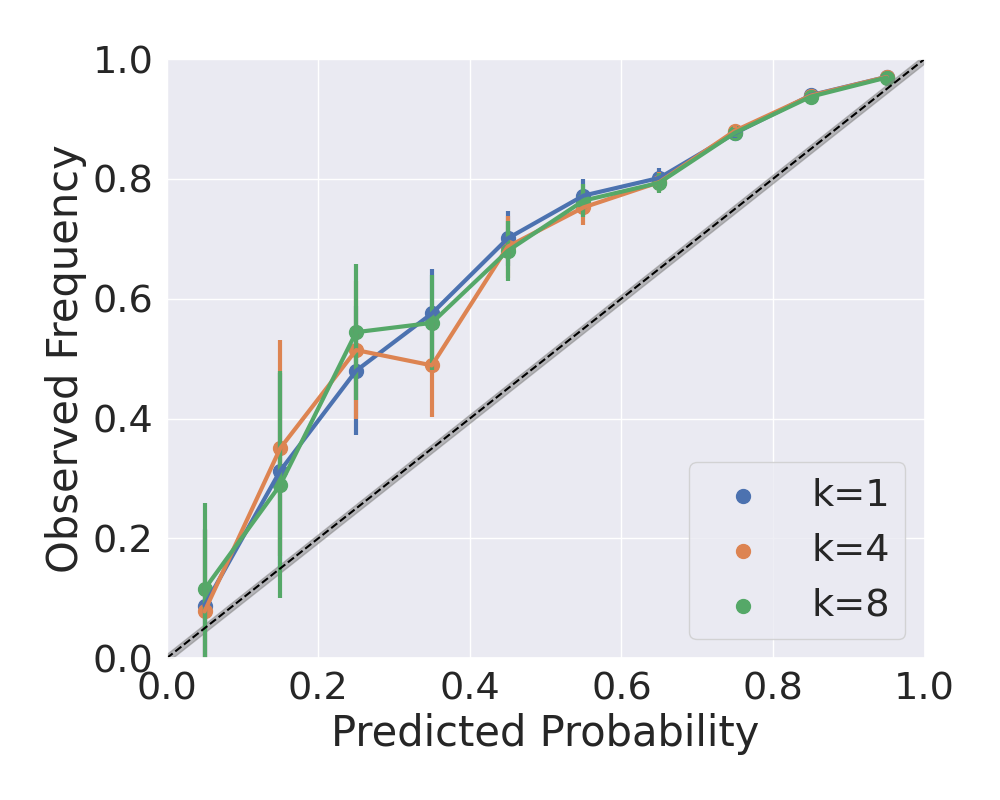}
  \caption{Household Group 1.}
  \label{fig:1}
\end{subfigure}%
\begin{subfigure}{.30\textwidth}
  \centering
  \includegraphics[width=\linewidth]{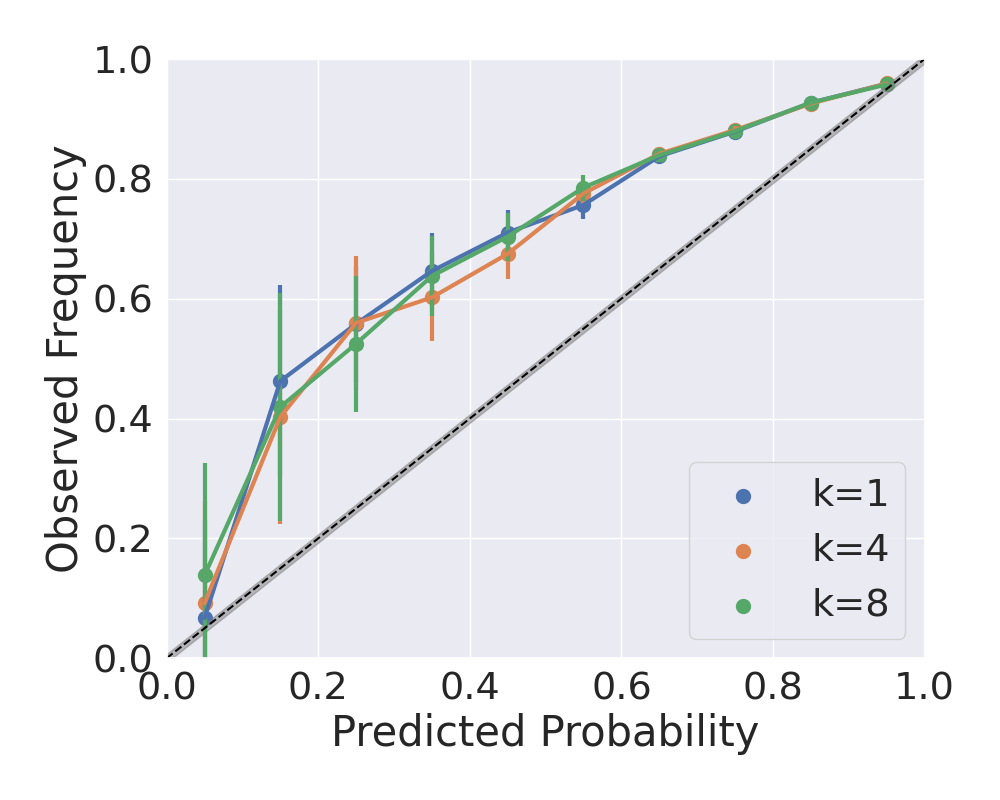}
  \caption{Household Group 2.}
  \label{fig:2}
\end{subfigure}
\begin{subfigure}{.30\textwidth}
  \centering
  \includegraphics[width=\linewidth]{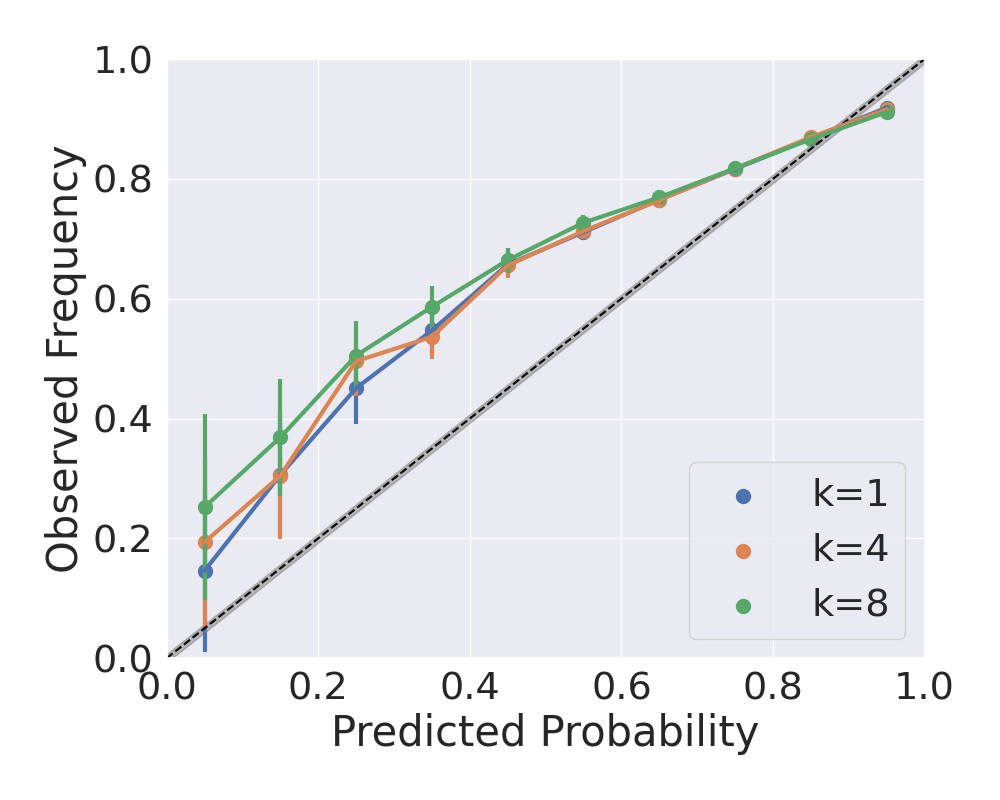}
  \caption{Household Group 3.}
  \label{fig:3}
\end{subfigure}
\caption{Multi-step ahead forecast calibration for a Bernoulli DGLM modeling p(Return) for household (a) Group 1, (b) Group 2 and (c) Group 3.  The forecasts are well calibrated across all forecast horizons considered. $k$ represents the forecast horizon in weeks.}
\label{fig:pReturn}
\end{figure}

To model whether a household returns to purchase any item in a given week, we model p(Return) with a Bernoulli DGLM, with a trend term and the covariate log total spend across all items for the previous week.  We also compare multi-step ahead forecasts at this modeling level, with forecast horizons of $k=1$ (one-week ahead), $k = 4$, (one month ahead) and $k = 8$ (two months ahead).  As we are modeling a binary outcome, we primarily focus on calibration plots to evaluate the forecasts, and find the calibration to be good for all household groups and across all three forecast horizons (\autoref{fig:pReturn}).  Additionally, we consider point forecasts under mean squared error, and the area under the curve and F1 score (the harmonic mean of the precision and recall), evaluating the binary forecasts as a binary classification problem.  Across all three of these metrics, the forecast accuracy is persistent across forecast horizons, with little drop in accuracy even for 2 month ahead forecasts ($k = 8$) (\autoref{tab:pReturn}).  

\begin{table}
\caption{AUC, F1 scores and MSE values for multi-step ahead forecasts with a Bernoulli DGLM fit to return to store.  All metrics are evaluated using the mean forecasts for each household for each week. $k$ represents the forecast horizon in weeks.}
\begin{center}
\vspace{-0.5cm}\scalebox{0.85}{
\begin{tabular}{c|l|c|c|c}
HH Group & Metric & $k = 1$ & $k = 4$ & $k = 8$ \\\hline
\multirow{ 3}{*}{1} & AUC & 0.68 &	0.68 &	0.67 \\
& F1 Score & 0.96 &	0.96 & 	0.95\\
& MSE & 0.07	& 0.07	& 0.08 \\\hline
\multirow{ 3}{*}{2} & AUC & 0.64 &	0.64 &	0.63\\
& F1 Score & 0.95 &	0.95 &	0.95\\
& MSE & 0.08 &	0.08	& 0.08\\\hline
\multirow{ 3}{*}{3} & AUC &  0.63 &	0.62	&0.62\\
& F1 Score & 0.90	&0.90 &	0.89\\
& MSE & 0.14 &	0.14 &	0.15\\
\end{tabular}}
\end{center}
\label{tab:pReturn}
\end{table}%

\subsection{Global Log Total Spend}

\begin{figure}
\centering
\begin{subfigure}{.30\textwidth}
  \centering
  \includegraphics[width=\linewidth]{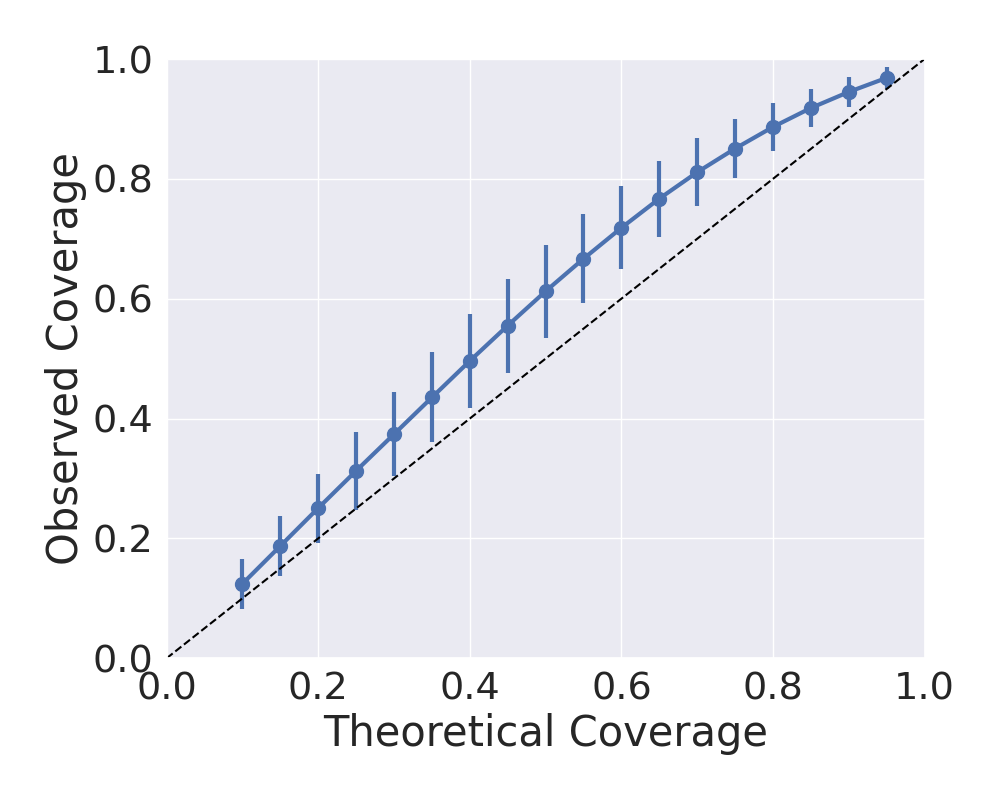}
  \caption{Household Group 1.}
  \label{fig:1}
\end{subfigure}%
\begin{subfigure}{.30\textwidth}
  \centering
  \includegraphics[width=\linewidth]{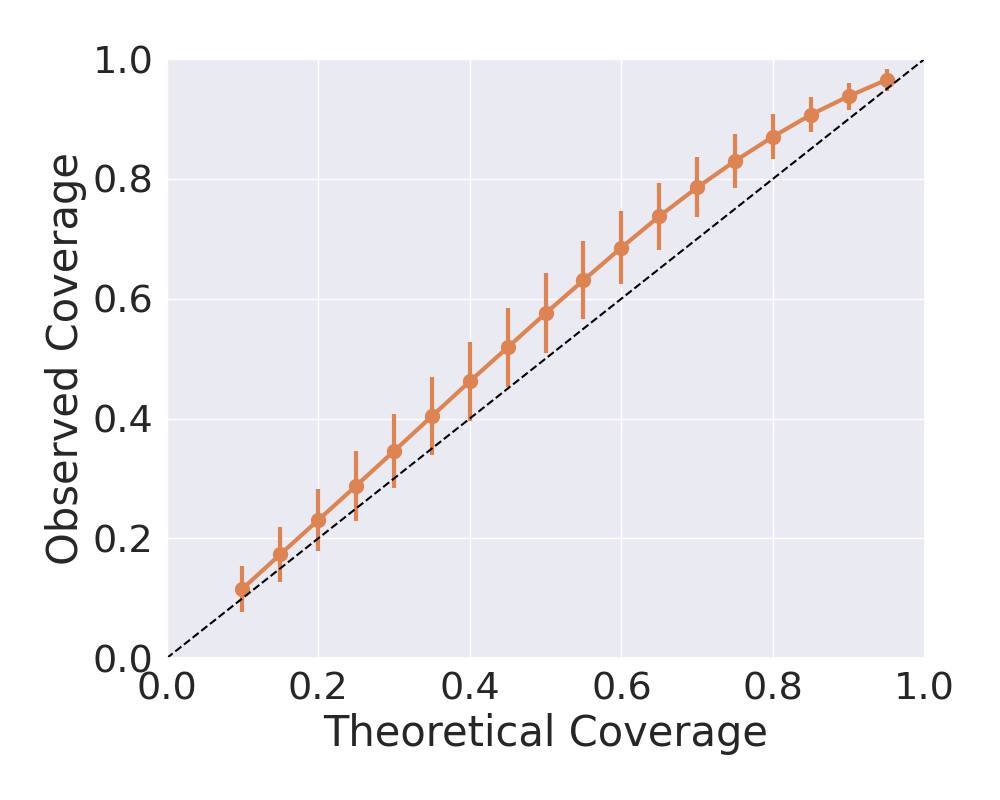}
  \caption{Household Group 2.}
  \label{fig:2}
\end{subfigure}
\begin{subfigure}{.30\textwidth}
  \centering
  \includegraphics[width=\linewidth]{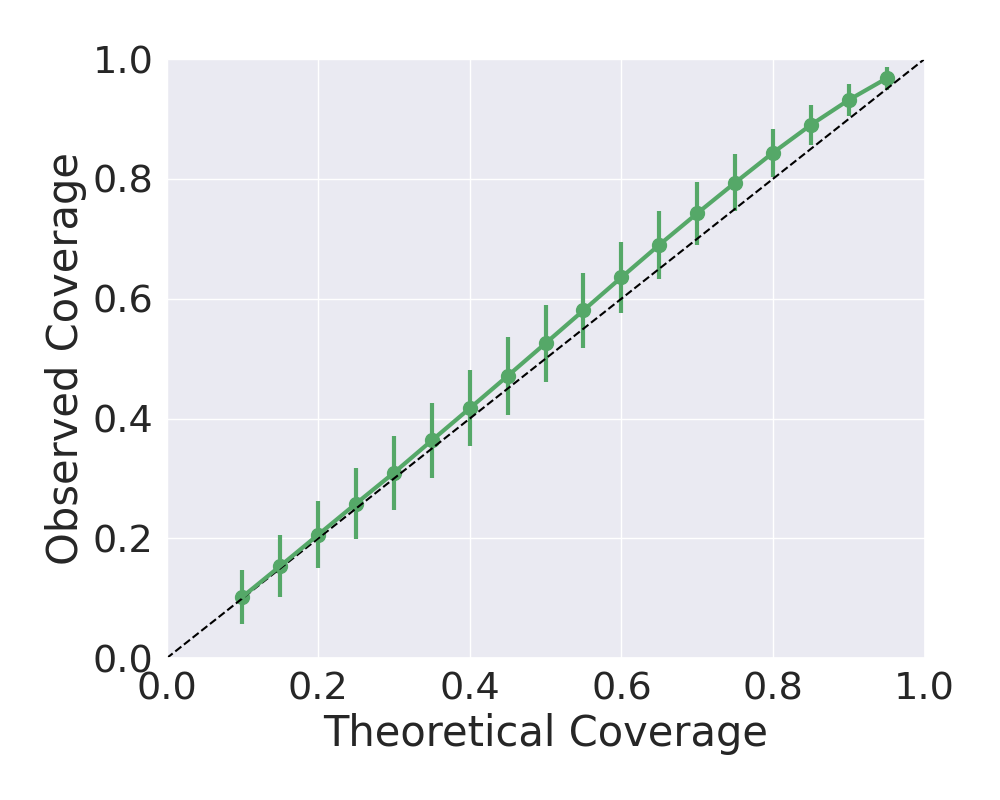}
  \caption{Household Group 3.}
  \label{fig:3}
\end{subfigure}
\caption{Coverage plots for a DLM modeling p(Global log Total Spend $\vert$ Return) for household (a) Group 1, (b) Group 2 and (c) Group 3 for one-step ahead forecasts.  The coverage plots evaluate the forecast uncertainty and indicate that the observed coverage aligns well with the expected theoretical coverage.}
\label{fig:log-total}
\end{figure}

At the next level in the modeling decomposition,   p(Global log Total Spend $\vert$ Return) is modeled using a DLM with a trend term and the covariate global log total spend at the last return.  These DLMs exhibit good coverage across all three household groups (\autoref{fig:log-total}), indicating that the uncertainty associated with these one-step ahead forecasts is well-calibrated.

\subsection{Category Level}

At the category level, we model p(log Total Spend by Category $\vert$ Return, Global log Total Spend) with a DLMM.  We compare models with three different predictors to evaluate the utility of the simultaneous predictors as compared to lagged predictors, where all covariates are treated as known.  All models have a trend term and an additional dynamic predictor, which is the log total spend in the category at the last return for M1 (a lagged, local predictor), the global log total spend across all categories at the last return (a lagged predictor) for M2 and the global log total spend (across all categories) for the \textit{current} week (simultaneous predictor) for M3.  Aggregate results for all three household groups are given for two additional categories in \autoref{tab:grp2-4} and \autoref{tab:grp2-8}. The results presented in the main paper were fit to the category containing Items A - D.  For each different metric, the point forecast is the optimal forecast under that loss function and MAPE is only evaluated at observations that are non-zero.   

\begin{table}[h]
\caption{Median forecast accuracy across all households in each household group  for DLMMs fit at the category level for the category containing Item E.  The metrics in parentheses represent the 25th and 75th percentile values across households.}
\begin{center}
\vspace{-0.5cm}\scalebox{0.85}{
\begin{tabular}{c|l|c|c|c}
HH Group & Metric & M1: Lagged Cat. & M2: Lagged Global & M3: Simultaneous Global \\\hline
\multirow{ 3}{*}{1} & MAD & 3.08, (1.96, 4.70)  & 3.04, (1.97, 4.50)  & $\bm{2.70}$, (1.83, 3.64) \\
& MAPE & 0.47, (0.39, 0.58) & 0.47, (0.39, 0.55)  & $\bm{0.40}$, (0.32, 0.48)  \\
& ZAPE  & 0.57, (0.48, 0.68) & 0.57, (0.47, 0.65) & $\bm{0.46}$, (0.35, 0.55) \\\hline
\multirow{ 3}{*}{2} & MAD &  2.98, (1.84, 4.40) & 3.16, (2.10, 4.51)  & $\bm{3.10}$, (2.08, 4.51)  \\
& MAPE & 0.53, (0.44, 0.66) & 0.55, (0.45, 0.68) & 0.57, (0.44, 0.71)  \\
& ZAPE &0.68, (0.58, 0.93) & 0.71, (0.61, 1.01) & $\bm{0.67}$, (0.54, 0.95) \\\hline
\multirow{ 3}{*}{3} & MAD & 1.62, (0.82, 2.66)  & 1.54, (0.76, 2.50)  &  1.55, (0.81, 2.46) \\
& MAPE & 0.58, (0.45, 0.74)  & 0.55, (0.44, 0.70)  & $\bm{0.54}$, (0.42, 0.71)  \\
& ZAPE & 0.72, (0.59, 1.11) & 0.70, (0.58, 1.02) & $\bm{0.63}$, (0.52, 0.89)  \\
\end{tabular}}
\end{center}
\label{tab:grp2-4}
\end{table}%

\begin{table}
\caption{Median forecast accuracy across all households in each household group for DLMMs fit at the category level, for the category containing Item F.  }
\begin{center}
\vspace{-0.5cm}\scalebox{0.85}{
\begin{tabular}{c|l|c|c|c}
HH Group & Metric & M1: Lagged Cat. & M2: Lagged Global & M3: Simultaneous Global \\\hline
\multirow{ 3}{*}{1} & MAD & 2.56, (1.45, 4.03)  & 2.54, (1.45, 3.97)  & 2.62, (1.51, 4.04) \\
& MAPE & 0.51, (0.40, 0.66) & 0.51, (0.40, 0.66)  & 0.53, (0.42, 0.68)  \\
& ZAPE  & 0.68, (0.60, 0.87) & 0.68, (0.59, 0.89) & $\bm{0.67}$, (0.59, 0.85) \\\hline
\multirow{ 3}{*}{2} & MAD &  1.43, (0.70, 2.42) & 1.39, (0.67, 2.24)  & $\bm{1.37}$, (0.67, 2.11)  \\
& MAPE & 0.42, (0.34, 0.55) & 0.40, (0.33, 0.48) & $\bm{0.38}$, (0.31, 0.44)  \\
& ZAPE &0.60, (0.53, 0.72) & 0.58, (0.52, 0.67) & $\bm{0.52}$, (0.46, 0.57) \\\hline
\multirow{ 3}{*}{3} & MAD & 1.38, (0.59, 2.31)  & 0.95, (0.39, 1.68)  &  0.95, (0.39, 1.63) \\
& MAPE & 0.51, (0.38, 0.67)  & 0.42, (0.35, 0.56)  & $\bm{0.39}$, (0.32, 0.51)  \\
& ZAPE & 0.67, (0.59, 1.07) & 0.60, (0.53, 0.73) & $\bm{0.52}$, (0.46, 0.59)  \\
\end{tabular}}
\end{center}
\label{tab:grp2-8}
\end{table}%

\subsection{Sub-Category Level}

At the sub-category level, we again model p(log Total Spend by Sub-Category $\vert$ Return in Category, Category log Total Spend) with a DLMM.  At this level of modeling, simultaneous predictors again serve to improve the forecasting accuracy.  We compare three models, each with a trend term and the additional predictors of M1: log total spend at the last return at the sub-category level (lagged), M2: log total spend at the last return in the category level (lagged) and M3: log total spend for the \textit{current} week at the category level (simultaneous).  The aggregate point forecast results are given in \autoref{tab:grp1-51} for the sub-category containing Item C and \autoref{tab:grp1-66} for the sub-category containing Item F.  Again, at the sub-category level, the use of simultaneous predictors greatly improves the forecast accuracy. 

\begin{table}[htp]
\caption{Median forecast accuracy across all households in each household group for DLMMs fit at the sub-category level for the sub-category containing Item C.  }
\begin{center}
\vspace{-0.5cm}\scalebox{0.9}{
\begin{tabular}{c|l|c|c|c}
HH Group & Metric & M1: Lagged Sub-Cat. & M2: Lagged Cat. & M3: Simultaneous Cat. \\\hline
\multirow{ 3}{*}{1} & MAD & 0.69, (0.23, 1.43)  & 0.70, (0.23, 1.43)  & $\bm{0.58}$, (0.18, 1.17) \\
& MAPE & 0.47, (0.35, 0.68) & 0.46, (0.34, 0.67)  & $\bm{0.36}$, (0.27, 0.52)  \\
& ZAPE  & 0.60, (0.49, 0.86) & 0.59, (0.50, 0.85) & $\bm{0.47}$, (0.40, 0.52) \\\hline
\multirow{ 3}{*}{2} & MAD &  0.41, (0.14, 0.93) & 0.41, (0.14, 0.92)  & $\bm{0.40}$, (0.13, 0.89)  \\
& MAPE & 0.48, (0.32, 0.71) & 0.44, (0.30, 0.64) & $\bm{0.36}$, (0.27, 0.57)  \\
& ZAPE & 0.55, (0.41, 0.67) & 0.53, (0.41, 0.65) & $\bm{0.46}$, (0.36, 0.52) \\\hline
\multirow{ 3}{*}{3} & MAD & 0.43, (0.17, 0.90)  & 0.42, (0.17, 0.90)  &  $\bm{0.37}$, (0.14, 0.78) \\
& MAPE & 0.54, (0.37, 0.74)  & 0.55, (0.37, 0.75)  & $\bm{0.41}$, (0.28, 0.63)  \\
& ZAPE & 0.59, (0.48, 1.15) & 0.60, (0.47, 1.15) & $\bm{0.45}$, (0.33, 0.51)  \\
\end{tabular}}
\end{center}
\label{tab:grp1-51}
\end{table}%

\begin{table}
\caption{Median forecast accuracy across all households in each household group for DLMMs fit at the sub-category level for the sub-category containing Item F.  }
\begin{center}
\vspace{-0.5cm}\scalebox{0.9}{
\begin{tabular}{c|l|c|c|c}
HH Group & Metric & M1: Lagged Sub-Cat. & M2: Lagged Cat. & M3: Simultaneous Cat. \\\hline
\multirow{ 3}{*}{1} & MAD & 1.86, (1.23, 2.55)  & 1.87, (1.27, 2.55)  & $\bm{1.25}$, (0.72, 1.71) \\
& MAPE & 0.45, (0.33, 0.59) & 0.45, (0.33, 0.60)  & $\bm{0.28}$, (0.21, 0.36)  \\
& ZAPE  & 0.62, (0.51, 1.18) & 0.62, (0.51, 1.19) & $\bm{0.43}$, (0.32, 0.51) \\\hline
\multirow{ 3}{*}{2} & MAD &  1.82, (1.22, 2.49) & 1.82, (1.21, 2.48)  & $\bm{1.13}$, (0.63, 1.55)  \\
& MAPE & 0.48, (0.34, 0.61) & 0.48, (0.34, 0.62) & $\bm{0.20}$, (0.37, 0.43)  \\
& ZAPE & 0.63, (0.52, 1.46) & 0.63, (0.52, 1.46) & $\bm{0.43}$, (0.31, 0.51) \\\hline
\multirow{ 3}{*}{3} & MAD & 1.62, (1.01, 2.31)  & 1.62, (1.00, 2.30)  &  $\bm{0.99}$, (0.52, 1.44) \\
& MAPE & 0.49, (0.35, 0.62)  & 0.49, (0.35, 0.63)  & $\bm{0.29}$, (0.21, 0.41)  \\
& ZAPE & 0.62, (0.51, 1.48) & 0.62, (0.51, 1.48) & $\bm{0.44}$, (0.31, 0.52)  \\
\end{tabular}}
\end{center}
\label{tab:grp1-66}
\end{table}%


\subsection{Item Level}

\begin{figure}
\centering
\begin{subfigure}{.30\textwidth}
  \centering
  \includegraphics[width=\linewidth]{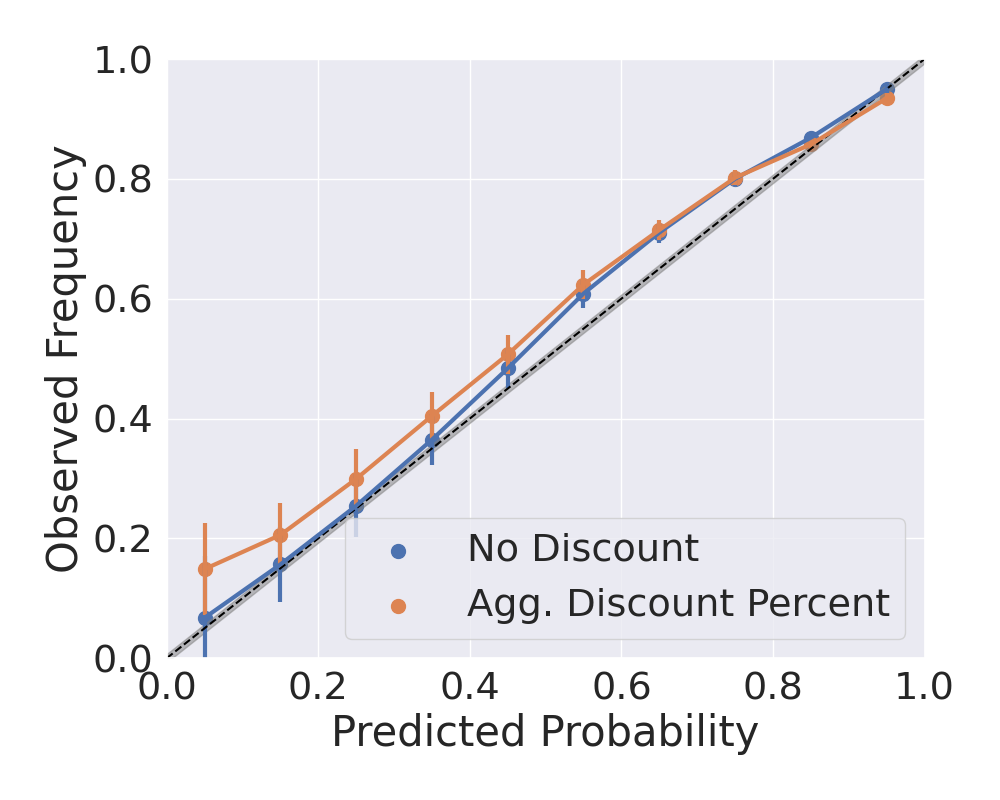}
  \caption{Item C.}
  \label{fig:1}
\end{subfigure}%
\begin{subfigure}{.30\textwidth}
  \centering
  \includegraphics[width=\linewidth]{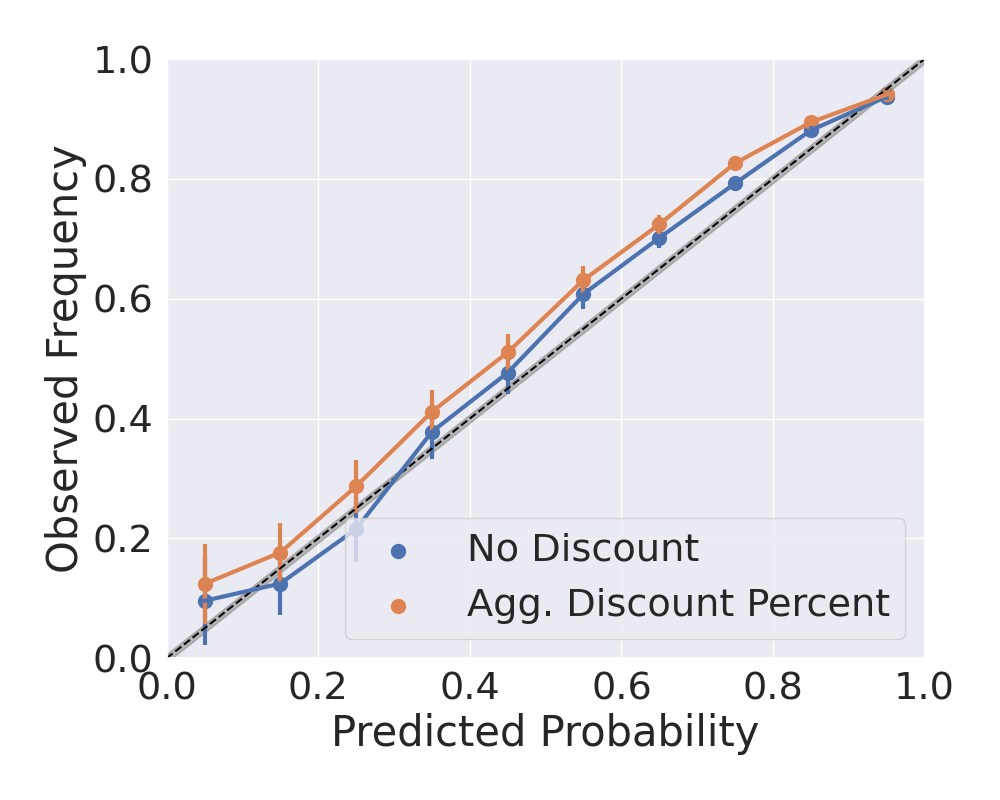}
  \caption{Item D.}
  \label{fig:2}
\end{subfigure}
\begin{subfigure}{.30\textwidth}
  \centering
  \includegraphics[width=\linewidth]{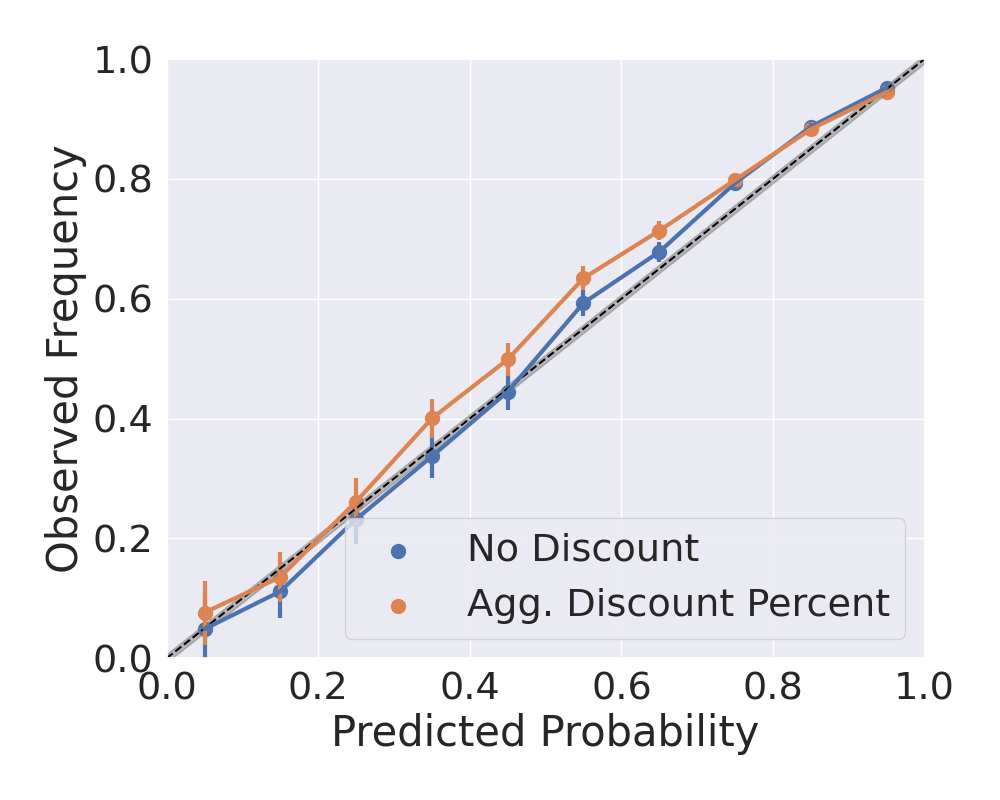}
  \caption{Item E.}
  \label{fig:3}
\end{subfigure}
\caption{Calibration plots for households in Group 1 on (a) Item C, (b) Item D, and (c) Item E for DLMMs with and without discount information.  Both models perform well across households in terms of calibration for all three items. }
\label{fig:calib-CDE}
\end{figure}

Calibration plots for DLMMs fit with and without discount information for Items C, D and E are presented for household Group 1 in \autoref{fig:calib-CDE}. Additional individual results for price sensitive households in household Group 2 (\autoref{fig:HH2-1-discount}) and household Group 3 (\autoref{fig:HH3-1-discount}) for Item A are presented below.  For each individual household, the inclusion of discount information improves forecasting accuracy and the state vector corresponding to the aggregate discount information is positive over time, representing the price sensitivity of each household.

\begin{figure}
\centering
\begin{subfigure}[b]{0.7\textwidth}
   \includegraphics[width=1\linewidth]{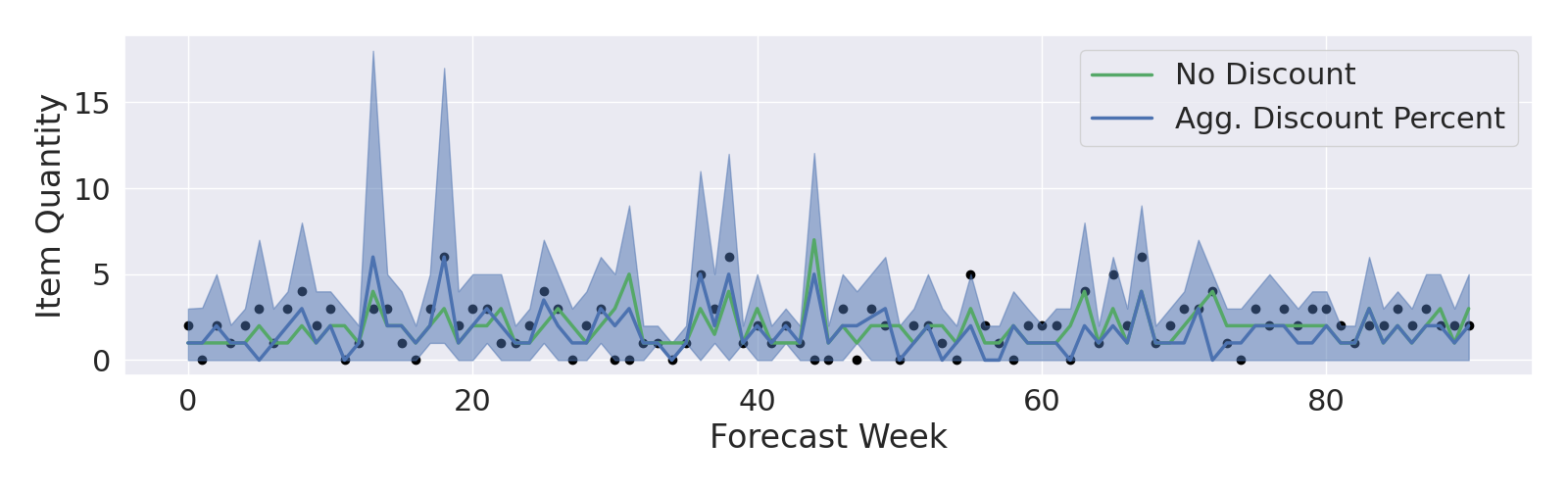}
   \caption{Forecasts}
   \label{fig:Ng1} 
\end{subfigure}

\begin{subfigure}[b]{0.7\textwidth}
   \includegraphics[width=1\linewidth]{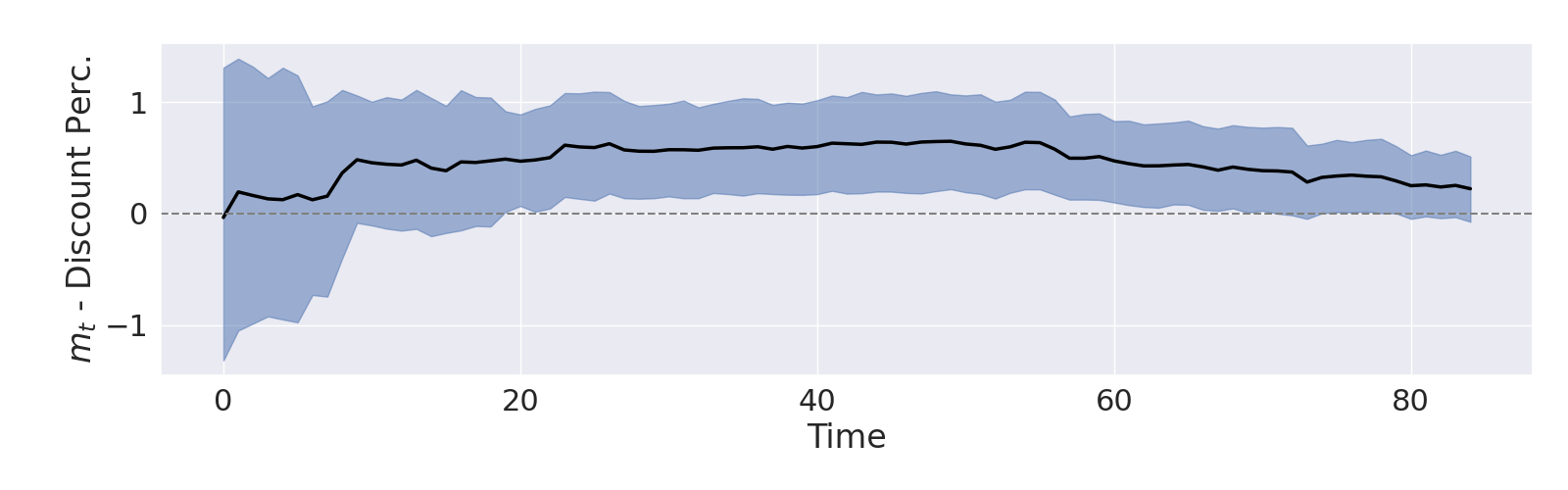}
   \caption{State Vector}
   \label{fig:Ng2}
\end{subfigure}
\caption{(a) MAD optimal point forecasts and  90\% prediction intervals for  Item A purchases of one price-sensitive household in Group 2, with and without discount information; (b) on-line posterior mean and  90\% intervals for the  state vector element  corresponding to the discount predictor.}\label{fig:HH2-1-discount}
\end{figure}

\begin{figure}
\centering
\begin{subfigure}[b]{0.7\textwidth}
   \includegraphics[width=1\linewidth]{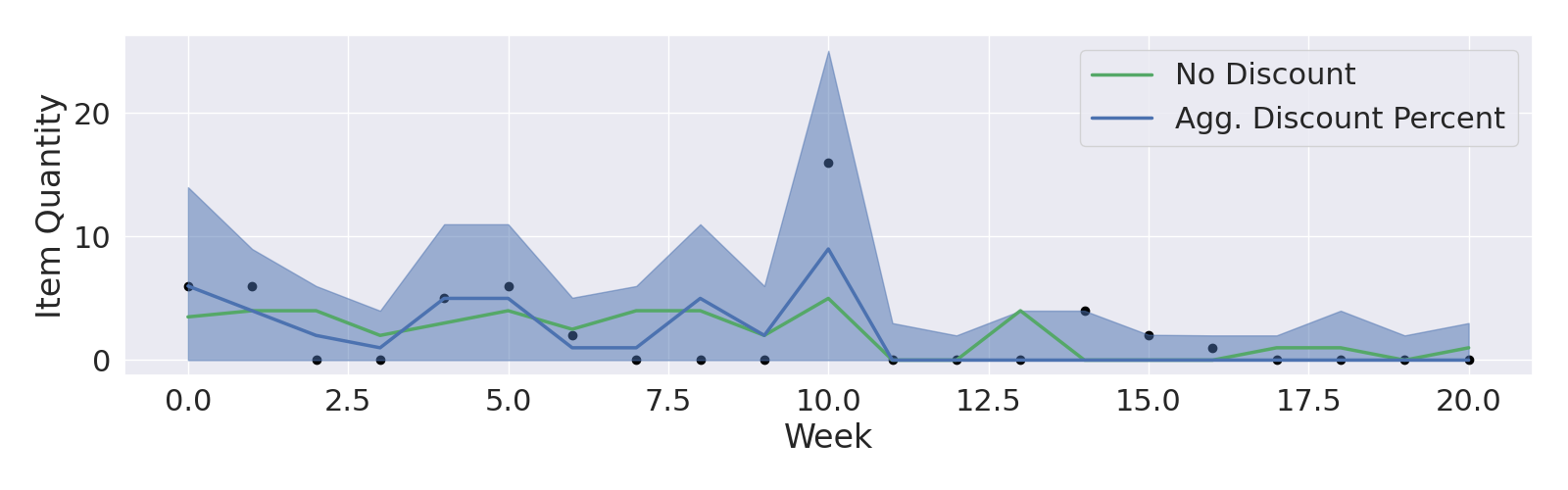}
   \caption{Forecasts}
   \label{fig:Ng1} 
\end{subfigure}

\begin{subfigure}[b]{0.7\textwidth}
   \includegraphics[width=1\linewidth]{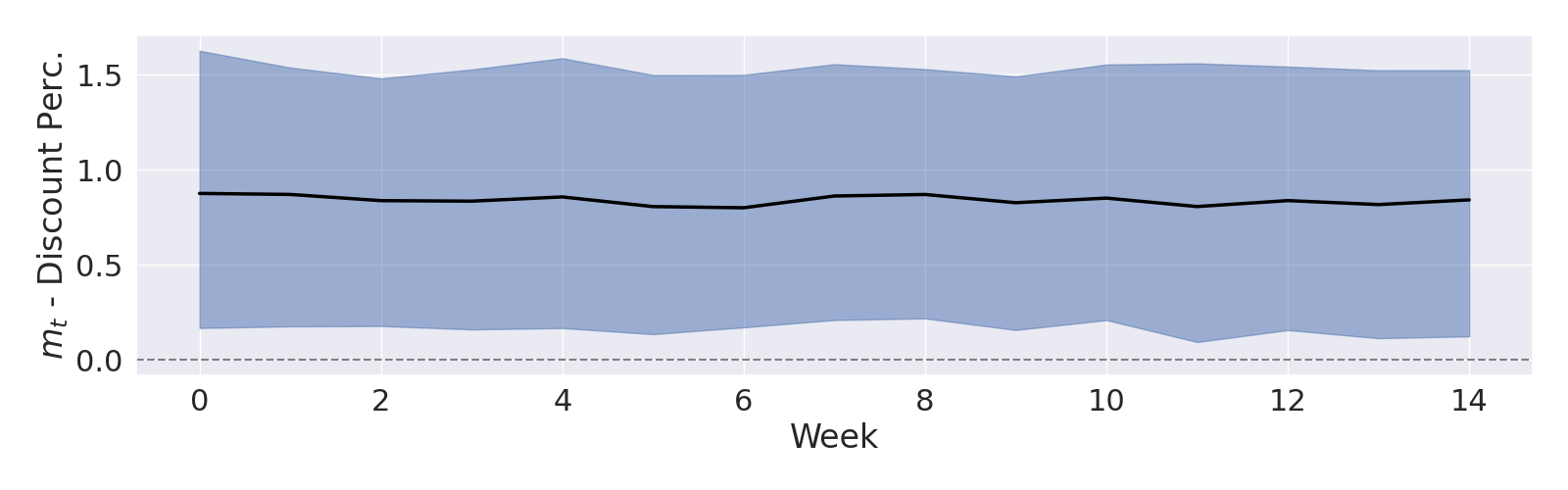}
   \caption{State Vector}
   \label{fig:Ng2}
\end{subfigure}
\caption{MAD optimal forecasts and other summaries for an individual household in Group 3, with format as in \autoref{fig:HH2-1-discount}.}\label{fig:HH3-1-discount}
\end{figure}

\end{document}